\documentclass[twocolumn,tighten]{aastex63}
\usepackage{graphicx}
\usepackage{amsmath}
\usepackage{scalerel}
\usepackage{apjfonts}
\usepackage{tikz}
\usetikzlibrary{svg.path}

\definecolor{orcidlogocol}{HTML}{A6CE39}
\tikzset{
  orcidlogo/.pic={
    \fill[orcidlogocol] svg{M256,128c0,70.7-57.3,128-128,128C57.3,256,0,198.7,0,128C0,57.3,57.3,0,128,0C198.7,0,256,57.3,256,128z};
    \fill[white] svg{M86.3,186.2H70.9V79.1h15.4v48.4V186.2z}
                 svg{M108.9,79.1h41.6c39.6,0,57,28.3,57,53.6c0,27.5-21.5,53.6-56.8,53.6h-41.8V79.1z M124.3,172.4h24.5c34.9,0,42.9-26.5,42.9-39.7c0-21.5-13.7-39.7-43.7-39.7h-23.7V172.4z}
                 svg{M88.7,56.8c0,5.5-4.5,10.1-10.1,10.1c-5.6,0-10.1-4.6-10.1-10.1c0-5.6,4.5-10.1,10.1-10.1C84.2,46.7,88.7,51.3,88.7,56.8z};
  }
}

\newcommand\orcidicon[1]{\href{https://orcid.org/#1}{\mbox{\scalerel*{
\begin{tikzpicture}[yscale=-1,transform shape]
\pic{orcidlogo};
\end{tikzpicture}
}{|}}}}

\usepackage{comment}
\usepackage{array,multirow,color,tablefootnote}
\usepackage{xcolor}

\begin{document}
\title{The Hubble PanCET Program: A Featureless Transmission Spectrum for WASP-29b and Evidence of Enhanced Atmospheric Metallicity on WASP-80b}

\author[0000-0001-9665-8429]{Ian~Wong \orcidicon{0000-0001-9665-8429}}
\altaffiliation{NASA Postdoctoral Program Fellow}
\affiliation{NASA Goddard Space Flight Center, 8800 Greenbelt Rd, Greenbelt, MD 20771, USA; \href{mailto:ian.wong@nasa.gov}{ian.wong@nasa.gov}}

\author[0000-0003-1728-8269]{Yayaati~Chachan \orcidicon{0000-0003-1728-8269}}
\affiliation{Division of Geological and Planetary Sciences, California Institute of Technology, 1200 East California Blvd, Pasadena, CA 91125, USA}

\author[0000-0002-5375-4725]{Heather~A.~Knutson \orcidicon{0000-0002-5375-4725}}
\affiliation{Division of Geological and Planetary Sciences, California Institute of Technology, 1200 East California Blvd, Pasadena, CA 91125, USA}

\author[0000-0003-4155-8513]{Gregory~W.~Henry \orcidicon{0000-0003-4155-8513}}
\affiliation{Center of Excellence in Information Systems, Tennessee State University, Nashville, TN 37209, USA}

\author[0000-0001-9897-9680]{Danica~Adams \orcidicon{0000-0001-9897-9680}}
\affiliation{Division of Geological and Planetary Sciences, California Institute of Technology, 1200 East California Blvd, Pasadena, CA 91125, USA}

\author[0000-0003-3759-9080]{Tiffany~Kataria \orcidicon{0000-0003-3759-9080}}
\affiliation{NASA Jet Propulsion Laboratory, 4800 Oak Grove Drive, Pasadena, CA 91109, USA}

\author[0000-0001-5578-1498]{Bj{\" o}rn~Benneke \orcidicon{0000-0001-5578-1498}}
\affiliation{Department of Physics and Institute for Research on Exoplanets, Universit{\' e} de Montr{\' e}al, Montr{\' e}al, QC, Canada}

\author[0000-0002-8518-9601]{Peter~Gao \orcidicon{0000-0002-8518-9601}}
\affiliation{Earth and Planets Laboratory, Carnegie Institute for Science, 5241 Broad Branch Road, NW, Washington, DC 20015, USA}

\author[0000-0001-5727-4094]{Drake~Deming \orcidicon{0000-0001-5727-4094}}
\affiliation{Department of Astronomy, University of Maryland, College Park, MD 20742, USA}

\author[0000-0003-3204-8183]{Mercedes L\'{o}pez-Morales \orcidicon{0000-0003-3204-8183}}
\affiliation{Center for Astrophysics ${\rm \mid}$ Harvard {\rm \&} Smithsonian, 60 Garden Street, Cambridge, MA 02138, USA}

\author[0000-0001-6050-7645]{David~K.~Sing \orcidicon{0000-0001-6050-7645}}
\affiliation{Department of Earth and Planetary Science, Johns Hopkins University, 3400 N. Charles Street, Baltimore, MD 21218, USA}
\affiliation{Department of Physics and Astronomy, Johns Hopkins University, 3400 N. Charles Street, Baltimore, MD 21218, USA}

\author[0000-0003-4157-832X]{Munazza~K.~Alam \orcidicon{0000-0003-4157-832X}}
\affiliation{Carnegie Earth \& Planets Laboratory, 5241 Broad Branch Road NW, Washington, DC 20015, USA}

\author[0000-0002-3891-7645]{Gilda~E.~Ballester \orcidicon{0000-0002-3891-7645}}
\affiliation{Lunar \& Planetary Laboratory, Department of Planetary Sciences, University of Arizona, Tucson, AZ 85721, USA}

\author[0000-0003-3726-5419]{Joanna~K.~Barstow \orcidicon{0000-0003-3726-5419}}
\affiliation{School of Physical Sciences, The Open University, Walton Hall, Milton Keynes MK7 6AA, UK}

\author[0000-0003-1605-5666]{Lars~A.~Buchhave \orcidicon{0000-0003-1605-5666}}
\affiliation{DTU Space, National Space Institute, Technical University of Denmark, Elektrovej 328, DK-2800 Kgs. Lyngby, Denmark}

\author[0000-0002-2248-3838]{Leonardo~A.~dos~Santos \orcidicon{0000-0002-2248-3838}}
\affiliation{Space Telescope Science Institute, 3700 San Martin Drive, Baltimore, MD 21218, USA}

\author[0000-0002-3263-2251]{Guangwei~Fu \orcidicon{0000-0002-3263-2251}}
\affiliation{Department of Astronomy, University of Maryland, College Park, MD 20742, USA}

\author[0000-0003-1756-4825]{Antonio~Garc\'ia~Mu\~noz \orcidicon{0000-0003-1756-4825}}
\affiliation{AIM, CEA, CNRS, Universit\'e Paris-Saclay, Universit\'e de Paris, F-91191 Gif-sur-Yvette, France}

\author[0000-0003-4816-3469]{Ryan~J.~MacDonald \orcidicon{0000-0003-4816-3469}}
\affiliation{Department of Astronomy and Carl Sagan Institute, Cornell University, 122 Sciences Drive, Ithaca, NY 14853, USA}

\author[0000-0001-5442-1300]{Thomas Mikal-Evans \orcidicon{0000-0001-5442-1300}}
\affiliation{Max Planck Institute for Astronomy, K{\"o}nigstuhl 17, D-69117 Heidelberg, Germany}

\author[0000-0002-1600-7835]{Jorge~Sanz-Forcada \orcidicon{0000-0002-1600-7835}}
\affiliation{Centro de Astrobiolog\'{i}a (CSIC-INTA), ESAC Campus, E-28692 Villanueva de la Ca\~nada, Madrid, Spain}

\author[0000-0003-4328-3867]{Hannah~R.~Wakeford \orcidicon{0000-0003-4328-3867}}
\affiliation{School of Physics, University of Bristol, H.~H.~Wills Physics Laboratory, Tyndall Avenue, Bristol BS8 1TL, UK}

\begin{abstract}
We present a uniform analysis of transit observations from the Hubble Space Telescope and Spitzer Space Telescope of two warm gas giants orbiting K-type stars---WASP-29b and WASP-80b. The transmission spectra, which span 0.4--5.0 $\mu$m, are interpreted using a suite of chemical equilibrium PLATON atmospheric retrievals. Both planets show evidence of significant aerosol opacity along the day--night terminator. The spectrum of WASP-29b is flat throughout the visible and near-infrared, suggesting the presence of condensate clouds extending to low pressures. The lack of spectral features hinders our ability to constrain the atmospheric metallicity and C/O ratio. In contrast, WASP-80b shows a discernible, albeit muted H$_2$O absorption feature at 1.4 $\mu$m, as well as a steep optical spectral slope that is caused by fine-particle aerosols and/or contamination from unocculted spots on the variable host star. WASP-80b joins the small number of gas-giant exoplanets that show evidence for enhanced atmospheric metallicity: the transmission spectrum is consistent with metallicities ranging from $\sim$30--100 times solar in the case of cloudy limbs to a few hundred times solar in the cloud-free scenario. In addition to the detection of water, we infer the presence of CO$_2$ in the atmosphere of WASP-80b based on the enhanced transit depth in the Spitzer 4.5 $\mu$m bandpass. From a complementary analysis of Spitzer secondary eclipses, we find that the dayside emission from WASP-29b and WASP-80b is consistent with brightness temperatures of $937 \pm 48$ and $851 \pm 14$ K, respectively, indicating relatively weak day--night heat transport and low Bond albedo. 
\end{abstract}

\section{Introduction}\label{sec:intro}

Transit spectroscopy has emerged as the leading method for studying exoplanet atmospheres in detail. By measuring the transit depth as a function of wavelength, this technique is capable of probing minute variations in atmospheric opacity along the day--night terminator, which reflect the spectral signatures of both gas-phase molecules and aerosol particles \citep[e.g.,][]{madhusudhan2014,crossfield2015,deming2017,kreidberg2018}. To date, well over 50 transmission spectra have been published, spanning a broad range of planet masses and equilibrium temperatures. These results have motivated population studies that seek to uncover overarching trends in the measured spectral properties, which may in turn elucidate the fundamental physical and chemical processes that shape planetary atmospheres across the galaxy. 

A major finding from these works is the ubiquity of aerosols in exoplanet atmospheres. Aerosols, i.e., small liquid or solid particles that are suspended in the atmosphere, encompass a diverse range of chemical compositions and physical characteristics. They are typically categorized into two broad groups: (1) condensate clouds, which form in equilibrium via phase changes and/or thermochemical reactions, and (2) photochemical hazes, which are created through direct interaction between the atmosphere and stellar irradiation (see the recent review by \citealt{gao2021}). The presence of aerosols attenuates molecular absorption features \citep[e.g.,][]{fortney2005,helling2008,morley2013,charnay2015,barstow2017,gao2018}, such as the water vapor feature at $\sim$1.4 $\mu$m. Meanwhile, the aerosol particle size distribution is reflected in visible wavelength transmission spectra: condensate clouds and aggregate photochemical haze particles larger than 1 $\mu$m contribute a flat (i.e., gray) opacity, while submicron hazes exhibit a characteristic scattering slope, with decreasing transit depth with increasing wavelength \citep[e.g.,][]{parmentier2013,barstow2017,lavvas2017,adams2019,powell2019,ohno2020b}. The impact of aerosols is especially apparent at equilibrium temperatures below $\sim$1000 K, where the majority of transmission spectra show heavily muted or null water absorption features in the near-infrared \citep[e.g.,][]{singstis,stevenson2016,crossfield2017,fu2017,tsiaras2018,gao2020,libbyroberts2020,dymont2021}, as well as a continuum of optical scattering slopes ranging from flat to super-Rayleigh \citep[e.g.,][]{heng2016only,singstis}.

In addition to their effects on the measured light from exoplanets, aerosols play a key role in shaping the composition and dynamics of planetary atmospheres. Opacity from aerosol particles can significantly alter the pressure--temperature profile on both local and planet-wide scales \citep[e.g.,][]{heng2012,morley2015,lavvas2021}. This interplay may in turn affect the relative abundances of key atmospheric molecules such as water and methane \citep[e.g.,][]{helling2019,molaverdikhani2020}. Increasingly sophisticated numerical modeling of aerosol formation has demonstrated the critical role played by a multitude of atmospheric properties, including metallicity, vertical mixing, and longitudinal transport \citep[e.g.,][]{parmentier2013,heng2015,helling2016,lee2016,lavvas2017,lines1,lines2,zhang2018,powell2019,woitke2020,christie2021}. However, the relationships between these properties and more fundamental characteristics of exoplanets are still largely unknown. With current and near-future telescope facilities offering increasingly detailed views into exoplanet atmospheres across the electromagnetic spectrum, a refined understanding of aerosols will be indispensable for adequately interpreting these new data.

In this paper, we examine the atmospheres of two warm gas giants---WASP-29b and WASP-80b---combining Hubble Space Telescope (HST) spectroscopic transit light curves in the optical and near-infrared with broadband Spitzer transit and secondary eclipse observations at 3.6 and 4.5 $\mu$m. WASP-29b is a Saturn-mass exoplanet ($M_p = 0.24\,M_{\rm Jup}$, $R_p = 0.78\,R_{\rm Jup}$) that orbits a 4800 K K3-type host star every 3.92 days \citep{hellier2010}. WASP-80b is larger, with a mass and radius of $0.54\,M_{\rm Jup}$ and $0.95\,R_{\rm Jup}$, and lies on a 3.07 day orbit around a 4145 K K6-type star \citep{triaud2013}. See Table~\ref{tab:properties} for a full summary of the target system properties. 

\begin{deluxetable}{lcccc}[t]
\tablewidth{0pc}
\setlength{\tabcolsep}{2pt}
\renewcommand{\arraystretch}{0.9}
\tabletypesize{\small}
\tablecaption{
    Target System Properties
    \label{tab:properties}
}
\tablehead{ &
    \multicolumn{1}{c}{WASP-29} &
    \multicolumn{1}{c}{Ref.}    &
    \multicolumn{1}{c}{WASP-80}  &
    \multicolumn{1}{c}{Ref.}   
}
\startdata
				 $T_{*}$ (K)  & $4800 \pm 150$& 1 & $4145 \pm 100$ & 4\\
				 $M_{*}$ ($M_{\odot}$)  & $0.825 \pm 0.033$ & 1 & $0.57 \pm 0.05$ & 4\\
				 $R_{*}$ ($R_{\odot}$)  & $0.808 \pm  0.044$ & 1 & $0.586^{+0.017}_{-0.018}$ & 5\\
				 Stellar $\log g$ &  $4.5 \pm 0.2$ & 1 & $4.663^{+0.015}_{-0.016}$ & 5 \\
				 Stellar [Fe/H] &  $+0.11 \pm 0.14$ & 1 & $-0.14 \pm 0.16$ & 4\\
				 $M_{p}$ ($M_{\rm Jup}$)  & $0.244 \pm 0.020$ & 1 & $0.538^{+0.035}_{-0.036}$ & 5\\
				 $R_{p}$ ($R_{\rm Jup}$)  & $0.776 \pm 0.043$ & 2 & $0.952^{+0.026}_{-0.027}$ & 4\\
				 $g_{p}$ (m~s$^{-2}$) & $9.6 \pm 2.7$ & 3 & $13.4 \pm 1.2$ & 3 \\
				 $P$ (d) & $3.9227183$ & 2 & $3.06786600$ & 6\\
				  & $\qquad\quad \pm 6.8$ e-06  & & $\qquad\quad \pm 3.6$ e-07 &\\
				 $a$ (au)  & $0.04565^{+0.00060}_{-0.00062}$ & 2 & $0.0346^{+0.0008}_{-0.0011}$ & 4\\
				 $i$ ($^{\circ}$) &  $89.17^{+0.50}_{-0.56}$ & 2 & $88.90 \pm 0.06$ & 7\\
		        $T_{\rm eq}$ (K)\tablenotemark{\scriptsize a}  & $970^{+32}_{-31}$ &  2 & $825 \pm 19$  & 5 \\
\enddata
\textbf{Note.}
\vspace{-0.2cm}\tablenotetext{\textrm{a}}{Planet dayside equilibrium temperature assuming zero Bond albedo and heat redistribution on the dayside hemisphere only.}
\textbf{References.} (1) \citealt{hellier2010}; (2) \citealt{gibson2013}; (3) \citealt{dymont2021}; (4) \citealt{triaud2013}; (5) \citealt{triaud2015}; (6) \citealt{parviainen2018}; (7) \citealt{sedaghati2017}.
\vspace{-0.8cm}
\end{deluxetable}

WASP-29b and WASP-80b join the growing population of warm ($500 < T_{\rm eq} < 1000$ K) gas giants with measured optical and near-infrared transmission spectra. Despite their disparate planet radii and masses, these two planets are similar with respect to both equilibrium temperature $T_{\rm eq}$ ($970^{+32}_{-31}$ vs. $825 \pm 19$ K) and surface gravity $\log\,g_{p}$ ($3.00 \pm 0.06$ vs. $3.17 \pm 0.04$; cgs units), making a comparative exploration of their atmospheric properties particularly apropos. Meanwhile, the corresponding host stars strongly differ in their activity levels, with WASP-80 showing significantly higher relative X-ray luminosity than WASP-29 (Section~\ref{subsec:xray}).

Previous ground-based visible spectroscopic transit light curves of WASP-29b revealed a flat spectrum, suggesting the absence of fine-particle aerosols \citep{gibson2013}. WASP-80b has been observed several times from the ground, with discrepant transmission spectrum shapes ranging from flat to steeply decreasing \citep{sedaghati2017,kirk2018,parviainen2018}. The HST near-infrared observations analyzed in this paper were previously fit by \citet{tsiaras2018}, who obtained a flat transmission spectrum for WASP-29b and a significant water-vapor absorption feature for WASP-80b. Our combined analysis of all HST and Spitzer observations provides an updated view of the planets' transmission spectra spanning 0.4--5.0 $\mu$m, with a uniform treatment of limb darkening, transit-shape modeling, and error analysis. We interpret the results of our light-curve fits using a suite of atmospheric retrievals, and derive constraints on atmospheric properties, with special attention given to exploring the characteristics of aerosols.

\section{Observations}\label{sec:obs}

We analyzed a total of six HST and Spitzer transit light curves for each target, along with a number of Spitzer secondary eclipse observations. A summary of all the datasets considered in this work is provided in Tables~\ref{hubbleobservations} and \ref{spitzerphotometry}.

\subsection{HST Transits}\label{subsec:hst}

Three transits of each system were observed with the Space Telescope Imaging Spectrograph (STIS) instrument in 2017 as part of the HST Panchromatic Comparative Exoplanetary Treasury Program (GO-14767; PIs: David Sing \& Mercedes L{\'o}pez-Morales). Two visits were conducted using the G430L grating (289--570 nm), and one visit was carried out with the G750L grating (526--1025 nm). The HST/STIS observations utilized the $52''\times2''$ slit and included wavelength calibration and flat-field exposures that were scheduled during the last spacecraft orbit of each visit. The per-exposure integration time was $t_{\rm int}=250$ s across all visits. Most visits consisted of five HST orbits and contained 48 individual exposures, with the exception of the second STIS G430L observation of WASP-80, for which only four HST orbits were allocated, yielding 38 exposures. We discarded the first of the two pre-ingress orbits during the light-curve extraction process (Section~\ref{subsec:stis}).

We observed one transit of each system in 2016 using the G141 grism (1.0--1.7 $\mu$m) on the Wide Field Camera 3 (WFC3) instrument as part of HST program GO-14260 (PI: Drake Deming). Both visits utilized the now-standard spatial scan mode in order to maximize the total flux from the target on the 266$\times$266 pixel subarray. For WASP-29, 74 exposures were obtained across five HST orbits, each with a total exposure time of 112 s. For WASP-80, the integration time was set to 103 s, and the 74 exposures were scheduled across four HST orbits. While the WASP-29 observations were conducted using only the forward scan direction, both forward and backward scans were included for WASP-80 in order to reduce overhead. A separate direct image of the target using the F139M grism was scheduled at the beginning of each visit, which was used to compute the wavelength solution of each exposure (Section~\ref{subsec:wfc3}).

\begin{deluxetable}{lcccc}[t]
\tablewidth{0pc}
\setlength{\tabcolsep}{3pt}
\renewcommand{\arraystretch}{0.9}
\tabletypesize{\small}
\tablecaption{
    HST Transit Observation and Data Reduction Details
    \label{hubbleobservations}
}
\tablehead{ \multicolumn{1}{c}{Dataset} &
    \multicolumn{1}{c}{UT Start Date} &
    \multicolumn{1}{c}{$n_{\rm exp}$\tablenotemark{\scriptsize a}}    &
    \multicolumn{1}{c}{$t_{\rm int}$\tablenotemark{\scriptsize b}}  &
    \multicolumn{1}{c}{Width\tablenotemark{\scriptsize c}}   
}
\startdata
				WASP-29 & & &  & \\
				\quad WFC3 G141 & 2016 Apr 15 & 74 & 112  & \dots \\
				\quad STIS G430L (1) & 2017 Jul 2 & 48 & 250  & 13\\
				\quad STIS G430L (2) & 2017 Jul 10 &  48 & 250  & 7\\	
				\quad STIS G750L & 2017 Aug 29 & 48 & 250 & 13\\
				\hline
				WASP-80 & & & & \\
				\quad WFC3 G141 & 2016 Jun 21 & 74 & 103  & \dots \\
				\quad STIS G430L (1) & 2017 May 15 & 48 & 250 & 11\\
				\quad STIS G430L (2) & 2017 Jul 13 &  38 & 250  & 9\\	
				\quad STIS G750L & 2017 Oct 25 & 48 & 250  & 15\\
\enddata
\textbf{Notes.}
\vspace{-0.2cm}\tablenotetext{\textrm{a}}{Number of exposures.}
\vspace{-0.2cm}\tablenotetext{\textrm{b}}{Per-exposure integration time, in seconds.}
\vspace{-0.2cm}\tablenotetext{\textrm{c}}{Width of the HST/STIS extraction aperture, in pixels.}
\vspace{-0.2cm}
\end{deluxetable}

\begin{deluxetable*}{lccccccccccc}[t]
\tablewidth{0pc}
\setlength{\tabcolsep}{8pt}
\renewcommand{\arraystretch}{0.9}
\tabletypesize{\footnotesize}
\tablecaption{
    Spitzer/IRAC Observation and Data Reduction Details
    \label{spitzerphotometry}
}
\tablehead{ & \vspace{-0.2cm}\\ \multicolumn{1}{c}{Dataset} & &
    \multicolumn{1}{c}{UT Start Date} & &
    \multicolumn{1}{c}{$n_{\rm img}$\tablenotemark{\scriptsize a}}    &
    \multicolumn{1}{c}{$t_{\rm int}$ (s)\tablenotemark{\scriptsize b}}  &
    \multicolumn{1}{c}{$t_{\rm trim}$ (min)\tablenotemark{\scriptsize c}}   & 
    \multicolumn{1}{c}{$r_{0}$\tablenotemark{\scriptsize c}} &  &
    \multicolumn{1}{c}{$r_{1}$\tablenotemark{\scriptsize c}} & 
    \multicolumn{1}{c}{$r_{\rm phot}$\tablenotemark{\scriptsize c}} & 
    \multicolumn{1}{c}{Binning\tablenotemark{\scriptsize d}}
}
\startdata
				WASP-29 & && && & & && & & \\
				3.6 $\mu$m & && && & & && & & \\
				\quad Transit & &2017 Feb 22 && 14016 & 1.92 & 45 & 3.0 & & 2.0  & $\sqrt{\beta}\times 1.35$ & 32 \\
				\quad Eclipse 1 && 2010 Aug 27 && 72512 & 0.36 & 60 & 2.5 && 2.0 & $\sqrt{\beta}\times 1.1$& 512\\
				\quad Eclipse 2 && 2011 Jan 11 && 79360 & 0.36 & 60 & 3.0 && 4.0 & $\sqrt{\beta}\times 1.25$& 256\\
				4.5 $\mu$m & &&& & & && & & & \\
				\quad Transit && 2017 Mar 14 && 14016 & 1.92 & 45 & 3.0 && \dots & 2.8 & 128 \\
				\quad Eclipse 1 && 2011 Jan 27  && 79360 & 0.36 & 30 & 4.0 && 1.5 & $\sqrt{\beta}\times0.95$ & 32 \\
				\quad Eclipse 2 && 2014 Aug 29 && 13952 & 1.92 & 0 & 3.5 && 1.5  & $\sqrt{\beta}+1.1$ & 128 \\
				\hline
				WASP-80 & && && & & && & & \\
				3.6 $\mu$m & && && & & && & & \\
				\quad Transit & &2013 Aug 1 && 40448 & 0.36 & 0 & 2.5 & & 1.5  & $\sqrt{\beta}\times1.2$ &512 \\
				\quad Eclipse 1 && 2013 Jul 3 && 40448 & 0.36 & 0 & 4.0 && \dots & 2.3 & 512\\
				\quad Eclipse 2 && 2013 Jul 24 && 40448 & 0.36 & 45 & 2.5 && 1.5 & $\sqrt{\beta}\times 1.15$& 512\\
				\quad Eclipse 3 && 2014 Jul 27 && 37696 & 0.36 & 45 & 4.0 && \dots & 1.9 & 512\\
				\quad Eclipse 4 && 2014 Jul 9 && 37696 & 0.36 & 30 & 3.0 && 2.0 & $\sqrt{\beta}\times 0.85$& 64\\
				4.5 $\mu$m & &&& & & && & & & \\
				\quad Transit && 2013 Jul 13 && 8576 & 1.92 & 30 & 3.0 & & 3.0 & $\sqrt{\beta}+1.4$ & 128 \\
				\quad Eclipse 1 && 2013 Jul 18 && 8576 & 1.92 & 0 & 2.5 && 3.0 & $\sqrt{\beta}\times1.5$ & 128 \\
				\quad Eclipse 2 && 2013 Jul 17 && 8576 &1.92 & 30 & 3.0 && 2.0  & $\sqrt{\beta}\times1.3$ & 64\\
\enddata
\textbf{Notes.}
\vspace{-0.2cm}\tablenotetext{\textrm{a}}{Number of images.}
\vspace{-0.2cm}\tablenotetext{\textrm{b}}{Integration time per image.}
\vspace{-0.2cm}\tablenotetext{\textrm{c}}{$t_{\rm trim}$ is the time interval trimmed from the start of the time series, $r_{0}$ is the aperture radius (in pixels) used for determining the star's centroid position, and $r_{1}$ is the aperture radius (in pixels) used for computing the noise pixel parameter $\beta$. The extraction aperture is defined in the $r_{\rm phot}$ column. See text for more details.}
\vspace{-0.2cm}\tablenotetext{\textrm{d}}{Bin size (in points) used on each photometric series prior to fitting.}
\vspace{-1cm}
\end{deluxetable*}

\subsection{Spitzer Transits and Secondary Eclipses}\label{subsec:spitzerobs}
A pair of transit observations in the 3.6 and 4.5 $\mu$m bandpasses of the Infrared Array Camera (IRAC) were obtained for each system. All of the Spitzer/IRAC observations utilized the 32$\times$32 pixel subarray mode. For WASP-29, the transit observations were conducted in 2017 as part of Program 13044 (PI: Drake Deming) and had an effective per-exposure integration time of 1.92 s. Short $\sim$30 minute peak-up observations were scheduled immediately before the start of the science exposures, to allow for the telescope pointing to stabilize. The 3.6 and 4.5 $\mu$m WASP-80b transit observations were carried out in 2013 as part of Program 90159 (PI: Amaury Triaud). For this program, the integration times in the two bandpasses were set to 0.36 and 1.92 s, respectively, and no peak-up observations were included.

Two 3.6 $\mu$m and one 4.5 $\mu$m occultation observations of WASP-29 were obtained in 2010 and 2011 (Programs 60003 and 70084; PI: Joseph Harrington). These visits used an integration time of 0.36 s and included no peak-up exposures. An additional 4.5 $\mu$m secondary eclipse observation with peak-up and $t_{\rm int}=1.92$ s was carried out in 2014 (Program 10054: PI: Heather Knutson). For WASP-80, two secondary eclipses were observed in each of the two IRAC bandpasses in 2013 as part of the same Spitzer program that provided the transit observations for the system. An additional pair of 3.6 $\mu$m occultation observations with peak-up and $t_{\rm int}=0.36$ s were conducted as part of Program 10054.

The Level 1 data products, which include dark-subtracted, flat-fielded, linearized, and flux-calibrated Spitzer/IRAC data processed by the official IRAC pipeline, were downloaded from the Spitzer Heritage Archive \citep{sha}.

\subsection{Stellar Activity}\label{subsec:activity}

An exoplanet's transmission spectrum can be affected by chromospheric activity on the host star; in particular, the slope of the transmission spectrum in the optical and near-infrared is susceptible to biases from unocculted starspots or faculae \citep[e.g.,][]{pont2008,sing2011,bruno2018,bruno2020,rackham2018,rackham2019}. The former can mimic the negative slope caused by Rayleigh scattering, while the latter tend to have the opposite effect. Long-term photometric monitoring is an important tool for probing the level of stellar variability and assessing the extent to which any stellar surface heterogeneities may affect the interpretation of the transmission spectra.

We regularly observed WASP-80 over 16 semesters from 2013 to 2021 using the Tennessee State University Celestron 14 inch Automated Imaging Telescope (AIT) at Fairborn Observatory in Arizona, USA. The AIT uses an SBIG STL-1001E CCD camera with a 1024$\times$1024 Kodak KAF-1001E CCD detector ($1\farcs2$ pixel scale). Images were obtained through a Cousins $R$-band filter. The original SBIG camera failed during the 2017B semester and was replaced with an identical camera and detector set; any observations from 2017B were not included in our photometric analysis. We calculated differential magnitudes of WASP-80 using the mean brightnesses of five nearby constant companion stars across 5--10 consecutive exposures. For more details about the data processing and analysis methodology, see \citet{sing2015} and \citet{wong2020}.

\begin{figure}[t]
\begin{center}
\includegraphics[width=\linewidth]{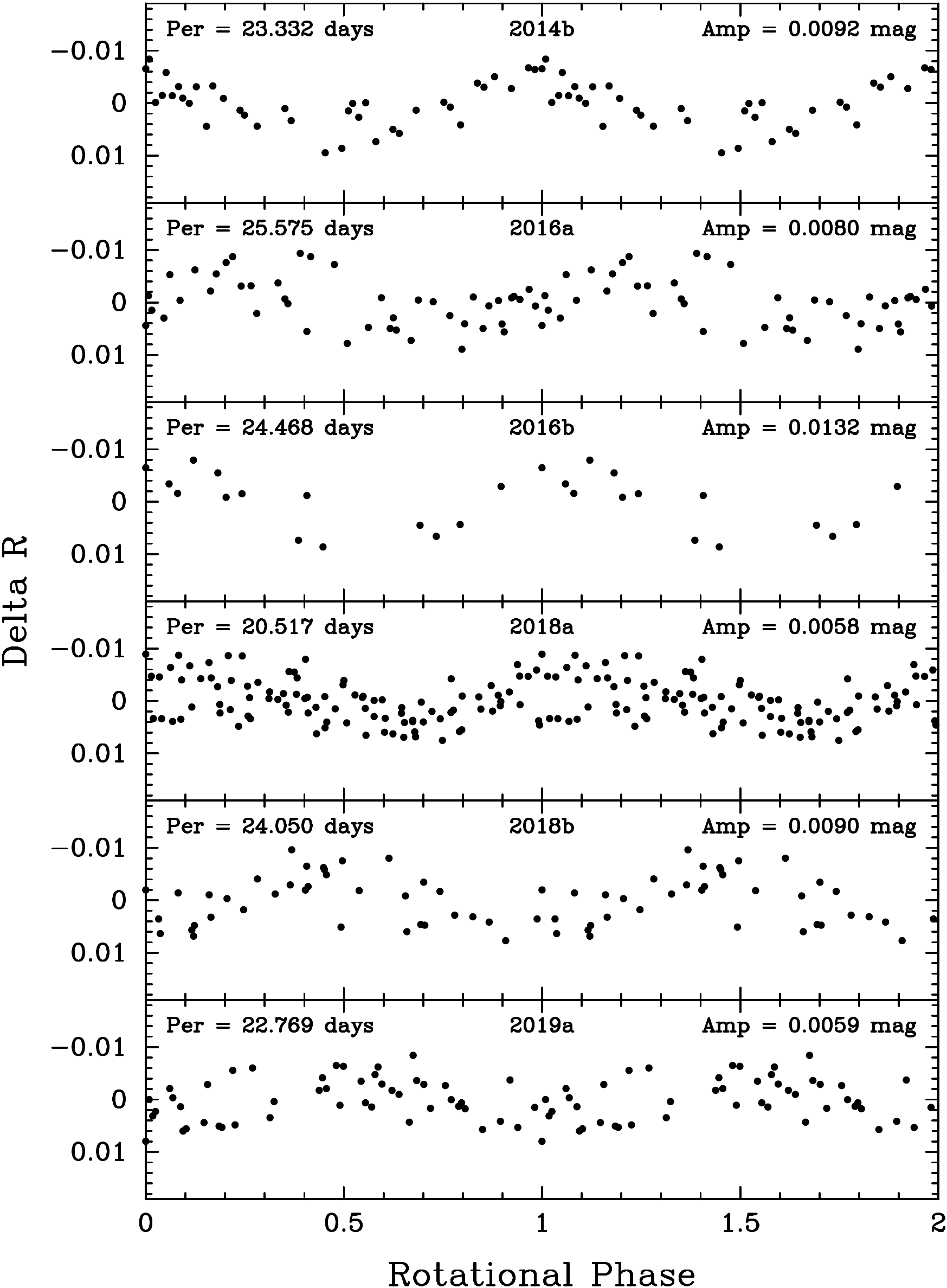}
\end{center}
\caption{Differential photometry of WASP-80 from our long-term photometric monitoring campaign at the AIT. Data are shown for the six semesters during which significant periodicity was detected. Each panel displays the relative $R$-band magnitude, phase-folded on the best-fit variability period. The period and peak-to-peak amplitude are listed for each semester.} \label{wasp80act}
\end{figure}

After removing outliers due to poor observing conditions and observations that coincided with a transit, we have a total of 596 nightly photometric measurements of WASP-80, with a typical single-measurement precision of 3--5 mmag on nights with good observing conditions. To search for stellar variability, we computed the Lomb--Scargle periodogram of each semester's photometric time series. Data from six of the 16 semesters (2014B, 2016A, 2016B, 2018A, 2018B, and 2019A) showed statistically significant peaks in the periodogram. The most prominent peaks corresponded to variability periods ranging from 20.52 to 25.58 days across the six semesters, with a mean photometric period of 23.45 days. The peak-to-peak brightness modulation of these periodic signals ranged from 5.8 to 13.2 mmag. Figure~\ref{wasp80act} shows the differential photometry of WASP-80 during the six semesters for which periodicity was detected; the light curves are phase-folded on the respective best-fit variability periods.

From their detailed statistical study of stellar activity in the Kepler sample, \citet{rackham2019} found that K6V stars such as WASP-80 have a $1\sigma$ range of peak-to-peak relative flux amplitudes spanning 0.35\%--1.28\%. Our measured $R$-band brightness modulation amplitudes for WASP-80 correspond to relative flux amplitudes of 0.54\%--1.22\%, entirely consistent with the reported range. Given the star's measured sky-projected rotational velocity and radius ($v\sin i_{*} = 1.27^{+0.14}_{-0.17}\,{\rm km}\,{\rm s}^{-1}$, $R_{*} = 0.586^{+0.017}_{-0.018}\,R_{\odot}$; \citealt{triaud2015}), the predicted rotation period is $P_{\rm rot}/\sin i_{*} = 23.3 \pm 3.2$ days, which is consistent with the average variability period we obtained from our photometric monitoring. We therefore attribute the source of the detected stellar activity to rotational brightness modulation from starspots, an interpretation that is also supported by the observed X-ray activity level (Section~\ref{subsec:xray}).

WASP-29 is located in the southern sky and is therefore inaccessible from the AIT. To place constraints on the level of stellar variability, we utilized the light curve provided by the Transiting Exoplanet Survey Satellite (TESS), which observed the target during Sector 2 (2018 August 22 to September 20) and Sector 29 (2020 August 26 to September 22). We considered the presearch conditioning simple aperture photometry (PDCSAP) light curve, which has been corrected by the official Science Processing and Operations Center (SPOC) pipeline \citep{jenkins2016} for common-mode systematics using cotrending basis vectors empirically derived from stars across the same detector array \citep{stumpe2014}. After trimming the planetary transits and removing $3\sigma$ outliers using a 16-point-wide moving median filter, we evaluated the scatter and photometric uncertainties at various binning intervals.

At the native 2 minute cadence, the scatter and median photometric error are 1150 and 1190 ppm for Sector 2 and 1110 and 1150 ppm for Sector 29, respectively. It follows that there is no significant short-term variability on minute-long timescales. Binning at one-hour intervals, we found that the binned scatter and median of the weighted mean photometric errors are 284 and 156 ppm for Sector 2 and 210 and 151 ppm for Sector 29, indicating a low level of time-correlated noise on timescales that are comparable to the transit duration. However, it is important to acknowledge the possible presence of uncorrected instrumental systematics trends, which have been shown to persist in the PDCSAP light curves, particularly in the vicinity of scheduled interruptions due to data downlinks and reaction wheel momentum dumps (see, for example, \citealt{wong2020year1,wong2021year2}). Visual inspection of the TESS WASP-29 light curve revealed flux ramps and discernible brightness discontinuities at several of the momentum dumps. Meanwhile, the Lomb--Scargle periodogram of each sector's light curve did not show any periodicities above $3\sigma$ significance. Our variability analysis of WASP-29 is insensitive to longer-term stellar variability ($>20$ d), given that each TESS sector spans less than a month. Therefore, while we clearly do not find brightness variations above a few 100s of ppm on timescales ranging from hours to days, we cannot confidently exclude the possibility that some lower-level stellar variability may exist. These results corroborate previous ultraviolet observations of WASP-29 that revealed a magnetically quiet star at short wavelengths \citep{dossantos2021}.

\subsection{X-ray Emission}\label{subsec:xray}

In order to further assess the activity level of the two host stars, we used archival data from the XMM-Newton observatory. WASP-80 was observed on 2014 May 13 and 2015 May 15 for a total of 49{,}000 s (formerly analyzed by \citealt{sal15} and \citealt{king18}). We reduced the data following standard procedures, removing intervals with high background levels. The filtered data from the two campaigns were combined, resulting in total exposure times of 39{,}300, 46{,}500, and 47{,}900 s in the pn, MOS1, and MOS2 detectors of the European Photon Imaging Camera (EPIC), respectively. We simultaneously fitted the spectra on the three detectors, using a two-temperature APED \citep[Astrophysics Plasma Emission Database;][]{aped} model with $\log \,T_{1,2} = 6.47\pm 0.08, 6.96\pm 0.08$ (in units of Kelvin), $\log\,{\rm EM}_{1,2} = 49.83^{+0.15}_{-0.21}, 49.70^{+0.15}_{-0.25}$ (in units of cm$^{-3}$),  [Fe/H]$=-0.14$, and an interstellar medium absorption of $\log\,N_{\rm H}=18.6$ (cm$^{-3}$), at a distance of 49.716 pc (Gaia EDR3).

The spectral fit yielded an X-ray luminosity of $L_{\rm X} = 3.24 \times 10^{27}$ erg\,s$^{-1}$, with a signal-to-noise ratio of 15.6 in the spectral range 5--100 \AA. We obtained a ratio of X-ray to bolometric luminosity of $\log L_{\rm X}/L_{\rm bol} = -4.9$, which indicates that WASP-80 has a moderate level of activity. A small flare was observed at X-ray wavelengths during the 2015 campaign, although no ultraviolet counterpart of the flare was detected by the optical monitor on board XMM-Newton. If we utilize the X-ray luminosity versus rotation period relation in \citet{wri11}, a rotation period of 23.45 days (Section \ref{subsec:activity}) would correspond to $L_{\rm X} = 6 \times 10^{27}$, consistent with the observed $L_{\rm X}$ value considering the level of variance in this relation.


Meanwhile, data for WASP-29 from XMM-Newton were presented and analyzed in \citet{dossantos2021}, where only an upper limit of $\log L_{\rm X}/L_{\rm bol} < -6.0$ could be derived, clearly indicating a significantly lower activity level than WASP-80. 

\section{Light-curve Extraction}\label{sec:lc}
We used the Python-based Exoplanet Transits, Eclipses, and Phase Curves (\texttt{ExoTEP}) pipeline to carry out the data processing and model fitting \citep[e.g.,][]{benneke2017, benneke2019}. The methodology we employed in our analysis is largely identical to the one described in \citet{wong2020}. In the following, we give a brief overview of the light-curve extraction process for each instrument.

\subsection{HST/STIS}\label{subsec:stis}

\begin{figure*}[t]
\begin{center}
\includegraphics[width=\linewidth]{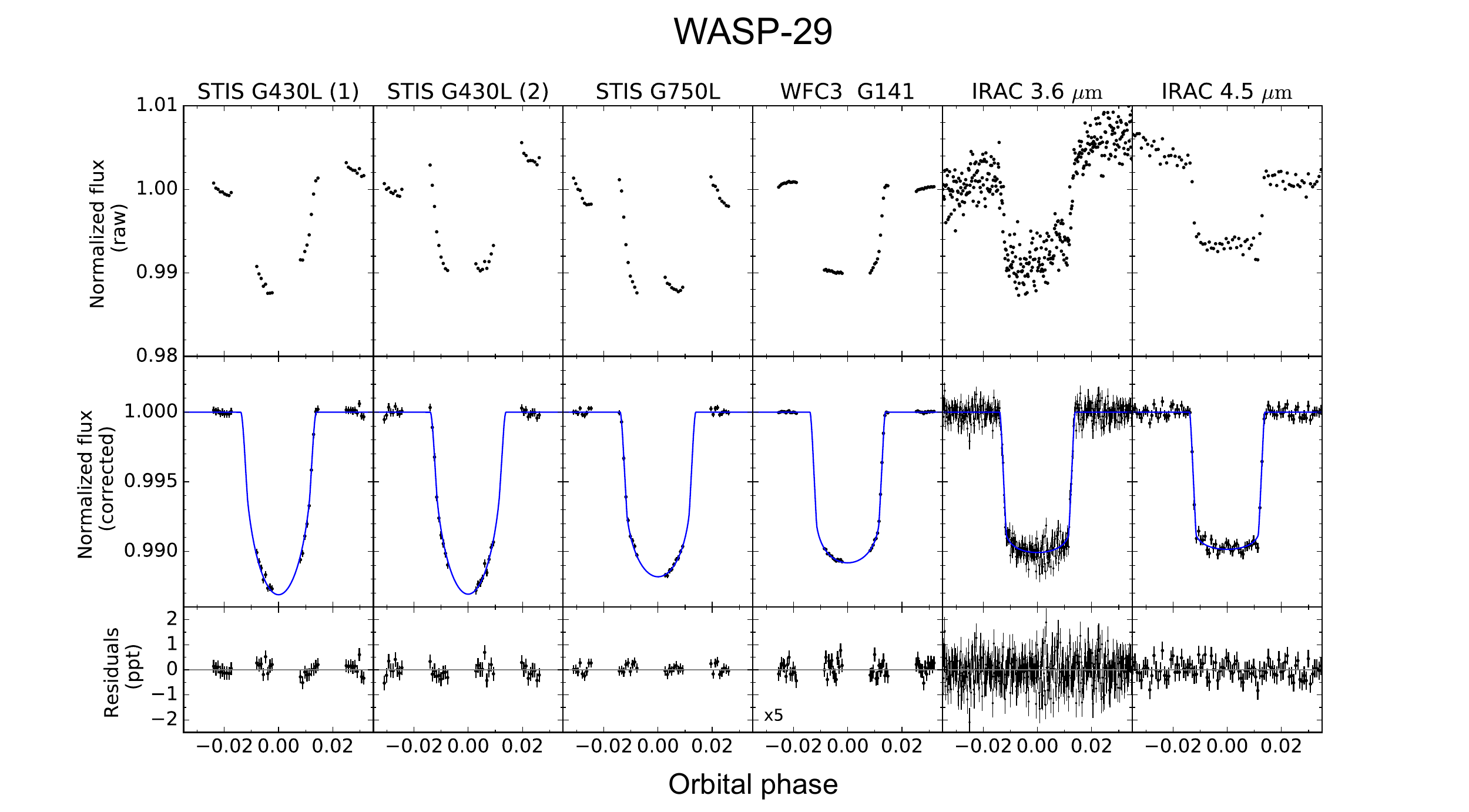}
\end{center}
\caption{Top panel: raw broadband light curves of all six WASP-29b transits considered in this work. The Spitzer light curves are binned by the optimized time intervals utilized in our fitting analysis. Middle panel: the corresponding systematics-corrected light curves, with the best-fit transit light curves shown in blue. Bottom panel: residuals from the best-fit instrumental and transit models. The error bars in each light curve are set to the corresponding best-fit photometric noise parameter. The WFC3 G141 residuals are amplified by five times for clarity.} \label{wasp29wlc}
\end{figure*}

\begin{figure*}[t]
\begin{center}
\includegraphics[width=\linewidth]{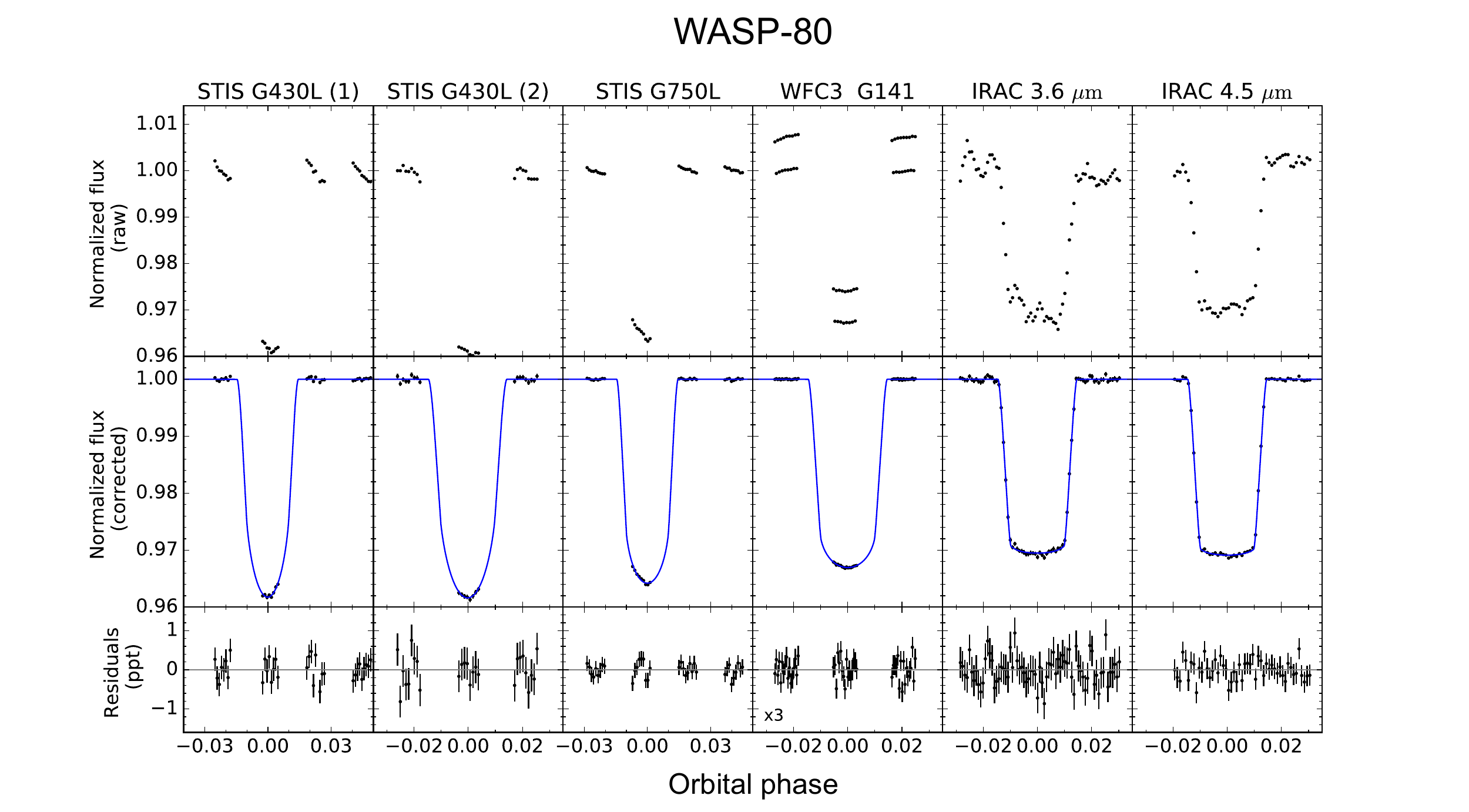}
\end{center}
\caption{Same as Figure~\ref{wasp29wlc}, but for WASP-80. The offsets in the WFC3 G141 raw light curve are due to the systematic flux differences between the forward and backward scans.} \label{wasp80wlc}
\end{figure*}

After downloading the raw images from the Mikulski Archive for Space Telescopes, we used the latest version of CALSTIS to flat-field them. We then replaced $4\sigma$ outlier pixel values in each individual frame with the corresponding median pixel values across all frames. The column-added 1D spectra were extracted from the background-subtracted frames using the wavelength solution provided by the accompanying calibrated \texttt{*sx1.fits} files. The width of the extraction aperture was chosen so as to minimize the scatter in the resultant residuals from the transit light-curve fit \citep[e.g.,][]{deming2013}. The optimal aperture widths for each visit are listed in Table~\ref{hubbleobservations}. For data obtained with the G750L grating, we corrected for the fringing effect using the fringe flat frame that was scheduled at the end of each visit \citep[see, for example,][]{nikolov,nikolov2,singstis}. 

To correct for small subpixel wavelength shifts across each visit, we set the first spectrum in the time series as the template and calculated the best-fit wavelength offsets and amplitude scaling factors using least-squares minimization to align each subsequent spectrum with the template. The normalized time series of best-fit amplitude scaling factors provided the broadband light curve for each visit. To extract the spectroscopic light curves, we used bins 100 pixels wide across the entire wavelength range spanned by the grating. To probe for the presence of alkali (Na and K) absorption lines, we included additional narrow wavelength bins at 588--591 and 766--772 nm for the G750L datasets. For these and all other light curves in our analysis, the time stamp of each spectrum was set to the mid-exposure time, converted to the BJD$_{\rm TDB}$ time standard \citep{eastman2010}

Following standard procedure, we discarded the first orbit prior to fitting. In addition, we removed the first data point from each orbit. Any remaining outliers in the light curves were removed by fitting each light curve separately and discarding points that were situated more than $5\sigma$ from the best-fit model. The 20th exposure was removed from the first STIS G430L visit of WASP-29. For the first STIS G430L transit observation of WASP-80, we discarded the 18th, 20th, and 35th spectra; the 20th exposure was trimmed from the second STIS G430L light curve. The raw outlier-trimmed HST/STIS broadband transit light curves of WASP-29 and WASP-80 used in our analysis are plotted in the top panels of Figures~\ref{wasp29wlc} and \ref{wasp80wlc}, respectively.

\subsection{HST/WFC3}\label{subsec:wfc3}

The HST/WFC3 G141 light curves were extracted from the dark- and bias-corrected \texttt{*ima.fits} files generated by CALWFC3. Following the usual technique for processing spatial scan mode images \citep[e.g.,][]{deming2013,kreidberg2014,evans2016}, we co-added the background-subtracted difference images from each pair of consecutive nondestructive reads to construct the full frame image. The top and bottom boundaries of each subexposure were determined from the row-added flux profile and included a buffer of 15 rows in both directions beyond the locations at which the flux falls to 20\% of its maximum value. We set the background flux level for each difference image as the median pixel value across two rectangular regions on either side of the spectral trace, making sure to avoid the edges of the subarray. The precise number of buffer rows and the locations of the background flux regions did not have any significant effect on the results of the light-curve fits. We corrected for cosmic rays by iteratively applying a 10 pixel wide moving median filter in the $x$ and $y$ directions and replacing $5\sigma$ outliers with the corresponding value in the row-added flux template, scaled to match the total column flux.

As described in detail in \citet{wong2020}, we followed the method developed by \citet{tsiaras} to compute the full 2D wavelength solution across the subarray. We first measured the relative horizontal offsets $\Delta x$ of each frame by comparing the centroid of the exposure with that of the first frame in the time series. Next, we utilized the calibration coefficients from the configuration file \texttt{WFC3.IR.G141.V2.5.conf} \citep{kuntschner} and the $x$-position of the star in the accompanying direct image to calculate the location of the spectral trace for a range of stellar $y$-positions and populate the wavelength solution at all pixels in the frame, correcting the $x$-position of the star in each frame by the previously determined offset. This wavelength solution was then used to flatten the full frame images using the coefficients provided in the \texttt{WFC3.IR.G141.flat.2.fits} calibration file \citep{kuntschner2}. 

To extract the spectroscopic light curves, we defined a 30 nm wavelength grid from 1.0 to 1.7 $\mu$m and used the wavelength solution to determine the precise locations of the wavelength boundaries across each frame. The total flux in each spectroscopic bin was obtained by adding up the pixel values within the corresponding region on the subarray. To accurately account for the partial pixels near the boundaries, we calculated the local 2D cubic interpolation function around each partial pixel and integrated it over the subpixel region to determine the fractional pixel value contained within the bin. The broadband light curve for each visit was obtained by adding up all the flux within the extraction aperture for each full frame image. We discarded the first orbit, as well as the first two data points from each subsequent orbit, prior to fitting. The raw HST/WFC3 transit light curves of WASP-29 and WASP-80 are included in the top panels of Figures~\ref{wasp29wlc} and \ref{wasp80wlc}, respectively.

\subsection{Spitzer IRAC}\label{subsec:spitzer}

The methodology we utilized to extract the IRAC light curves closely mirrors the techniques described in several previous analyses of post-cryogenic Spitzer datasets \citep[e.g.,][]{knutson2012,lewis,todorov,wong2,wong3,wong2020}. We determined the background level in each flux-calibrated subarray by masking out the star and the diffraction spikes, as well as the aberrant top row, and then calculating the median from a Gaussian fit to the pixel value distribution. After subtracting the background, we iteratively removed $3\sigma$ outliers across the time series at each pixel position using a moving median filter spanning 64 adjacent images.

The star's position on the subarray was computed using flux-weighted centroiding \citep[e.g.,][]{knutson2012}. We also calculated the noise pixel parameter $\beta$ \citep[for details, see][]{lewis}, which quantifies the width of the target's point response function (PRF). We considered a range of circular apertures of radii $r_{0}$ and $r_{1}$ with which to measure the centroid position and $\beta$, respectively, yielding different versions of the photometric series. The final photometric extraction was carried out using both fixed and time-varying circular apertures of radii $r_{\rm phot}$ centered on the centroid position. The radii of the time-varying apertures were defined with respect to the square-root of the noise pixel parameter $\sqrt{\beta}$ via either a multiplicative scaling factor or a constant shift \citep{wong2,wong3,wong2020}. Outlier removal was performed on the photometric series, star centroid positions, and the noise pixel parameter array using a 3$\sigma$ moving median filter with a width of 64 data points.

In addition to altering $r_{0}$, $r_{1}$, and $r_{\rm phot}$, we binned the light curves prior to fitting, with bin sizes ranging from 1 to 512 points in powers of two. To remove possible residual flux ramps at the beginning of each time series, we also experimented with trimming the first 15, 30, 45, or 60 minutes of data, while ensuring that at least 30 minutes of pre-ingress photometry remained in the trimmed time series to adequately establish the out-of-eclipse baseline in the fits. For each visit, the optimal choice of the various apertures, bin size, and trimming interval was determined by fitting each version of the photometry to the full transit or secondary eclipse light-curve model (see Section~\ref{sec:analysis}) and choosing the dataset that produced the smallest scatter in the residuals, binned in five-minute intervals. The optimal photometric extraction parameters are listed in Table~\ref{spitzerphotometry} for each visit. The raw Spitzer/IRAC light curves of WASP-29 and WASP-80 are shown in the top panels of Figures~\ref{wasp29wlc} and \ref{wasp80wlc}.

\section{Analysis}\label{sec:analysis}
The full light-curve model used in our fits consists of the astrophysical transit or secondary eclipse model and various instrumental systematics models. These are described separately in the following, along with the multistep fitting procedure used to produce the final results.

\subsection{Transit and Secondary Eclipse Model}\label{subsec:model}

The \texttt{ExoTEP} pipeline generates the transit and secondary eclipse model light curves using \texttt{batman} \citep{batman}. In our global analysis of the broadband transit light curves, the transit depth for each visit (parameterized by the planet--star radius ratio $R_{p}/R_{*}$) was allowed to vary independently. We also fit for common orbital ephemeris (mid-transit time $T_{0}$ and orbital period $P$) and transit-shape parameters (impact parameter $b$ and scaled semimajor axis $a/R_{*}$), which are shared across all visits for each system. Both systems are consistent with circular orbits, and we fixed the orbital eccentricity to zero in all of our transit fits.

For the HST/STIS and HST/WFC3 datasets, the limb-darkening coefficients for both broadband and spectroscopic light curves were computed using the \texttt{LDTk} package \citep{ldtk}, which is incorporated into \texttt{ExoTEP}. From the input stellar parameters of the host star, this program calculates the mean limb-darkening profile for each bandpass from interpolated $50-2600$ nm PHOENIX stellar intensity profiles \citep{husser} and fits for the limb-darkening coefficients using maximum-likelihood optimization. Following \citet{wong2020}, we chose the standard four-parameter nonlinear limb-darkening law and fixed the limb-darkening coefficients to the values obtained from \texttt{LDTk}. In the case of the Spitzer transit light curves---for which \texttt{LDTk} is not applicable, due to the wavelength range of the PHOENIX models---we set the limb-darkening coefficients to the tabulated values\footnote{\url{pages.jh.edu/~dsing3/David_Sing/Limb_Darkening.html}} provided by \citet{sing} for the nearest available combination of stellar parameters.

We experimented with fitting for the quadratic limb-darkening coefficients, but the relatively poor precision of the Spitzer data and the incomplete ingress/egress coverage of the HST light curves did not provide good constraints on the coefficient values, resulting in significantly larger uncertainties on the transit depths. Nevertheless, when comparing the results from the light-curve fits with free and fixed limb darkening coefficients, we found that the astrophysical parameter values agreed with each other to within $1\sigma$ in all cases.

\subsection{Instrumental Systematics}\label{subsec:systematics}

Each raw light curve contains systematics trends that are characteristic of the respective instrument, which we modeled using standard detrending functions. 

We employed the prescription described in \citet{sing2008} as the basis for our HST/STIS systematics modeling. We considered three variants:

\begin{align}
S_{\rm STIS,lin}(t) &= (c+vt_v)\cdot F(t_{\rm orb}), \label{stissystematics1} \\
S_{\rm STIS,linx}(t) &= (c+vt_v)\cdot F(t_{\rm orb}) + m\Delta x, \label{stissystematics2} \\
S_{\rm STIS,quad}(t) &= (c+vt_v + qt_v^2)\cdot F(t_{\rm orb}), \label{stissystematics3} \\
F(t_{\rm orb}) &= 1+\sum\limits_{k=1}^{4} p_{k}t_{\rm orb}^k. \label{combination}
\end{align}
The parameters $c$ and $v$ are the normalization constant and the slope of the visit-long linear trend, respectively; $t_{v}$ is defined as the time since the start of the visit. The systematics trend within each spacecraft orbit $F(t_{\rm orb})$ is modeled as a fourth-order polynomial with coefficients $p_{1-4}$; $t_{\rm orb}$ is the time from the start of each orbit. The basic model \texttt{lin} includes just a visit-long linear trend. The \texttt{linx} version has an additional linear trend as a function of the horizontal offset $\Delta x$ of each spectrum relative to the first (see Section~\ref{subsec:stis}). An additional quadratic term in time is included in the \texttt{quad} variant. 

We determined the optimal choice of systematics model for each visit by minimizing the resultant scatter in the residuals as well as the Akaike Information Criterion \citep{aic}: ${\rm AIC}\equiv 2\gamma-2\log L$, where $\gamma$ is the number of free parameters, and $\log L$ is the log-probability of the best-fit model. For WASP-29, we found that the two G430L light curves and the G750L light curve prefer the \texttt{lin}, \texttt{lin}, and  \texttt{linx} detrending models, respectively. In the case of the three HST/STIS observations of WASP-80, we chose \texttt{quad}, \texttt{linx}, and  \texttt{quad}, respectively. The particular choice of systematics models did not have any significant effect on the best-fit astrophysical parameters. However, the optimal combination of models yielded more consistent transmission spectrum shapes between the two G430L visits than other combinations.

\begin{deluxetable*}{lcccc}[t]
\tablewidth{0pc}
\tabletypesize{\small}
\renewcommand{\arraystretch}{0.9}
\tablecaption{
    WASP-29b and WASP-80b Global Fit Results
    \label{tab:globalfit}
}
\tablehead{ & & & \underline{WASP-29b} & \underline{WASP-80b} \\
\multicolumn{1}{c}{Parameter} &
    \multicolumn{1}{c}{Instrument} &
    \multicolumn{1}{c}{Wavelength (nm)}    &
    \multicolumn{1}{c}{Value}  &
    \multicolumn{1}{c}{Value}   
}
\startdata
				Relative planet radius, $R_{p}/R_{*}$ & STIS G430L (1) & 289--570 & $0.09609 \pm 0.00047$ & $0.17239_{-0.00058}^{+0.00057}$ \\
				Relative planet radius, $R_{p}/R_{*}$ & STIS G430L (2) & 289--570 &  $0.09598^{+0.00047}_{-0.00050}$ &  $0.17311_{-0.00087}^{+0.00093}$ \\
				Relative planet radius, $R_{p}/R_{*}$ & STIS G750L & 526--1025 & $0.09706^{+0.00049}_{-0.00052}$ & $0.17391^{+0.00041}_{-0.00043}$ \\
				Relative planet radius, $R_{p}/R_{*}$ & WFC3 G141 & 920--1800 &  $0.09697^{+0.00015}_{-0.00014}$ &  $0.17247^{+0.00021}_{-0.00020}$\\
				Relative planet radius, $R_{p}/R_{*}$ & IRAC 3.6 $\mu$m & 3161--3928 & $0.09720^{+0.00037}_{-0.00038}$  & $0.17213^{+0.00038}_{-0.00039}$ \\
				Relative planet radius, $R_{p}/R_{*}$ &  IRAC 4.5 $\mu$m & 3974--5020 &  $0.09645 \pm 0.00050$ &  $0.17361^{+0.00036}_{-0.00035} $\\
				Transit center time, $T_{0}$  & & & &\\
				\quad (BJD$_{\rm TDB}-2{,}456{,}000$) & \dots& \dots & $1807.235305 \pm 0.000098$ & $505.832170 \pm 0.000044$ \\
				Period, $P$ (days) & \dots& \dots & $3.9227090_{-0.0000018}^{+0.0000017}$ & $3.0678569 \pm 0.0000016$ \\
				Impact parameter, $b$ & \dots& \dots & $0.092^{+0.059}_{-0.060}$ & $0.305^{+0.016}_{-0.018}$ \\
				Inclination,\tablenotemark{\scriptsize a} $i$ (deg) & \dots& \dots &  $89.58^{+0.027}_{-0.028}$ &  $88.596^{+0.089}_{-0.084}$\\
				Relative semimajor axis, $a/R_{*}$ & \dots & \dots &$12.512^{+0.050}_{-0.086}$  &$12.451^{+0.073}_{-0.071}$ \\
\enddata
\textbf{Note.}
\vspace{-0.2cm}\tablenotetext{\textrm{a}}{Inclination is derived from the impact parameter via $b=(a/R_{*})\cos i$.}
\end{deluxetable*}

For the HST/WFC3 broadband light curves, the systematics model was defined as follows \citep[e.g.,][]{deming2013,kreidberg2014}:

\begin{equation}\label{wfc3systematics}S_{\rm WFC3}(t) = (c \times s^* + vt_v) \cdot (1 - \exp[-at_{\rm orb} - b - D^*]).\end{equation}
Here, the parameters $c$, $v$, $t_{v}$, and $t_{\rm orb}$ are analogous to the corresponding parameters in Equations~\eqref{stissystematics1}--\eqref{combination}. The additional piecewise-defined parameter $s^{*}$ was included for the WASP-80 HST/WFC3 visit, where both forward and backward scans were used, in order to account for the common-mode offset between the forward-scan and backward-scan fluxes (see Figure~\ref{wasp80wlc}): it is set to 1 for the forward scans and $s$ for the backward scans. The second term models the exponential ramp that occurs within each spacecraft orbit: $a$ and $b$ are the rate constant and amplitude. The shape of the ramp in the first fitted orbit (after removing the initial orbit from the original time series; Section~\ref{subsec:wfc3}) differs from those in the subsequent orbits. This is accounted for by the piecewise-defined parameter $D^*$, which is set to $d$ for points in the first fitted orbit and $0$ everywhere else.

\begin{deluxetable}{lccc}[t!]
\setlength{\tabcolsep}{3pt}
\tablewidth{0pc}
\renewcommand{\arraystretch}{0.9}
\tabletypesize{\small}
\tablecaption{
    Spectroscopic Light-curve Fit Results
    \label{tab:specfit}
}
\tablehead{ & \underline{WASP-29b} & & \underline{WASP-80b} \\
\multicolumn{1}{l}{Wavelength (nm)} &
    \multicolumn{1}{c}{$R_{p}/R_{*}$} & &
    \multicolumn{1}{c}{$R_{p}/R_{*}$}   
}
\startdata
STIS G430L & & \\
401--428 & $0.09744 \pm 0.00103$ & & \dots \\
428--456 & $0.09692 \pm 0.00089$ & & $0.17427 \pm 0.00084$ \\
456--483 & $0.09660 \pm 0.00048$ & & $0.17352 \pm 0.00068$ \\
483--511 & $0.09645 \pm 0.00073$ & & $0.17360 \pm 0.00074$ \\
511--538 & $0.09646 \pm 0.00066$ & & $0.17298 \pm 0.00095$ \\
538--565 & $0.09601 \pm 0.00053$ & & $0.17334 \pm 0.00064$ \\
STIS G750L & & \\
528--577 & $0.09613 \pm 0.00113$ & & $0.17324 \pm 0.00069$ \\
577--626 & $0.09753 \pm 0.00109$ & & $0.17361 \pm 0.00047$ \\
626--674 & $0.09707 \pm 0.00081$ & & $0.17403 \pm 0.00063$ \\
674--723 & $0.09603 \pm 0.00081$ & & $0.17369 \pm 0.00040$ \\
723--772 & $0.09551 \pm 0.00124$ & & $0.17216 \pm 0.00043$ \\
772--821 & $0.09682 \pm 0.00102$ & & $0.17320 \pm 0.00050$ \\
821--870 & \dots & & $0.17332 \pm 0.00048$ \\
870--919 & \dots & & $0.17281 \pm 0.00044$ \\
919--967 & \dots & & $0.17219 \pm 0.00082$ \\
588--591\tablenotemark{\scriptsize a} & $0.0986 \pm 0.0034$ & & $0.1720 \pm 0.0023$ \\
766--772\tablenotemark{\scriptsize a} & $0.0976 \pm 0.0034$ & & $0.1740 \pm 0.0013$ \\
WFC3 G141 & & \\
1130--1160 & $0.09728 \pm 0.00036$ & & $0.17204 \pm 0.00026$ \\
1160--1190 & $0.09729 \pm 0.00034$ & & $0.17228 \pm 0.00030$ \\
1190--1220 & $0.09727 \pm 0.00036$ & & $0.17240 \pm 0.00027$ \\
1220--1250 & $0.09713 \pm 0.00034$ & & $0.17226 \pm 0.00024$ \\
1250--1280 & $0.09724 \pm 0.00030$ & & $0.17243 \pm 0.00026$ \\
1280--1310 & $0.09699 \pm 0.00035$ & & $0.17257 \pm 0.00019$ \\
1310--1340 & $0.09689 \pm 0.00037$ & & $0.17268 \pm 0.00027$ \\
1340--1370 & $0.09735 \pm 0.00032$ & & $0.17319 \pm 0.00023$ \\
1370--1400 & $0.09723 \pm 0.00023$ & & $0.17284 \pm 0.00024$ \\
1400--1430 & $0.09689 \pm 0.00034$ & & $0.17284 \pm 0.00029$ \\
1430--1460 & $0.09727 \pm 0.00032$ & & $0.17298 \pm 0.00022$ \\
1460--1490 & $0.09719 \pm 0.00032$ & & $0.17286 \pm 0.00025$ \\
1490--1520 & $0.09655 \pm 0.00034$ & & $0.17258 \pm 0.00033$ \\
1520--1550 & $0.09800 \pm 0.00041$ & & $0.17229 \pm 0.00026$ \\
1550--1580 & $0.09677 \pm 0.00031$ & & $0.17250 \pm 0.00029$ \\
1580--1610 & $0.09693 \pm 0.00037$ & & $0.17235 \pm 0.00024$ \\
1610--1640 & $0.09733 \pm 0.00037$ & & $0.17236 \pm 0.00024$ \\
\enddata
\textbf{Note.}
\vspace{-0.1cm}\tablenotetext{\textrm{a}}{Narrowband light curves centered on the Na and K absorption regions, respectively.}
(This table is available in its entirety in machine-readable form.)
\vspace{-0.2cm}
\end{deluxetable}

Spitzer IRAC photometry is affected by the well-known intrapixel sensitivity variations. In our analysis, we used Pixel Level Decorrelation  \citep{pld} to detrend these variations. The systematics model is based on a linear combination of the nine pixel-flux arrays $\hat{P}_{k}$ extracted from the $3\times 3$ box centered on the star:

\begin{equation}\label{pld}S_{\rm IRAC}(t) = 1+\sum\limits_{k=1}^{9}w_{k}\hat{P}_{k}(t_i)+ vt_v.\end{equation}
The coefficients $w_{k}$ are the weights on the respective pixel arrays. In addition, we included a visit-long linear trend with slope $v$. The pixel arrays were binned in the same manner as the photometric light curve prior to fitting (see Section~\ref{subsec:spitzer}).

\subsection{Fitting Procedure}\label{subsec:fitting}

\texttt{ExoTEP} simultaneously calculates the best-fit parameter values and uncertainties using the Markov Chain Monte Carlo (MCMC) package \texttt{emcee} \citep{emcee}. In addition to the astrophysical and systematics parameters, the pipeline includes the per-point uncertainties $\sigma_{i}$ for each light curve as additional photometric noise parameters. We allowed $\sigma_{i}$ to vary freely to ensure that the resultant reduced $\chi^2$ was unity and to self-consistently generate realistic error bars on the fitted parameters. 

The global analysis of each planet's transmission spectrum consisted of three stages. In the first stage, we carried out a joint fit to the HST/WFC3 and Spitzer/IRAC broadband light curves. The Spitzer photometry were prebinned by the intervals that we determined previously from optimizing the photometric extraction of each visit individually (Table~\ref{spitzerphotometry}). In this fit, we allowed all of the orbital ephemeris and transit-shape parameters to vary, along with the three transit depths in the respective bandpasses. 

This methodology differs from the one described in \citet{wong2020} in that we did not include the HST/STIS broadband light curves in the joint fit. Our choice to exclude those datasets at this stage was motivated by the low signal-to-noise ratio (when compared to the HST/WFC3 datasets) as well as the relatively sparse coverage of the transit ingress and egress, particularly in the case of WASP-80b. When we experimented with including all broadband light curves in the joint fit, we found that the transit-shape parameters and mid-transit time shifted to better match some of the HST/STIS points during ingress and egress, at the expense of a poorer fit to the continuously sampled Spitzer transit light curves.

The second stage of our fitting analysis consisted of fitting the HST/STIS broadband light curves individually, with Gaussian priors placed on $b$, $a/R_{*}$, $T_{0}$, and $P$ using the values determined from the joint fit to the HST/WFC3 and Spitzer/IRAC light curves in the previous stage. For each planet, we analyzed the two STIS G430L visits separately in order to probe for possible stellar variability that may affect the transit depths across the different epochs. After this stage, the global broadband light-curve analysis was complete.

For the final stage, we measured the transmission spectra by fitting the spectroscopic light curves in the STIS G430L, STIS G750L, and WFC3 G141 bandpasses. Here, we allowed only $R_{p}/R_{*}$ to vary, while fixing the transit-shape parameters and orbital ephemeris to the best-fit values from the broadband light-curve fits. Following \citet{wong2020}, we included the full HST/STIS systematics model when fitting the individual spectroscopic light curves, given that the shape of the systematics trends varies appreciably across the bandpass. Meanwhile, for the WFC3 G141 spectroscopic light-curve fits, we first divided out a common-mode correction from the spectroscopic light curves using the ratio of the raw broadband flux array and the best-fit broadband transit model \citep[e.g.,][]{deming2013}. We then applied a simplified systematics model to account for the remaining instrumental flux variations in each spectroscopic light curve:

\begin{equation}\label{specsystematics}S_{\rm spec}(t) = c+v\cdot{\rm\Delta}x.\end{equation}
This two-parameter model consists of a normalization constant $c$ and a linear slope $v$ with respect to the measured subpixel shift $\Delta x$ of each spectrum in the time series relative to the first one (Section~\ref{subsec:wfc3}).

Using the precorrection and simplified systematics modeling for the WFC3 G141 spectroscopic light curves takes advantage of the more uniform shape of the systematics trends in the near-infrared bandpass, reducing the number of free parameters and yielding tighter constraints on the transit depths. No significant improvement to the residual scatter or AIC was found when applying the full HST/WFC3 systematics model to the spectroscopic light curves.

\section{Results}\label{sec:results}

\subsection{Broadband Light-curve Fits}\label{subsec:wlcfits}

The results of our global broadband light-curve analyses of WASP-29 and WASP-80 are listed in Table~\ref{tab:globalfit}. The systematics-corrected light curves and best-fit transit models are plotted in Figures~\ref{wasp29wlc} and \ref{wasp80wlc}. Notably, when comparing the transit depths from the two STIS G430L visits, we find that the values agree with one another to within $1\sigma$ for both planets, indicating the absence of significant stellar photometric variability between the two epochs.

Comparing our best-fit transit shape parameters for WASP-29b with the results from \citet{gibson2013}, we find that our values for $b$ and $a/R_{*}$ are consistent with theirs at better than the $1\sigma$ level, while having significantly smaller uncertainties. Meanwhile, we obtained an updated orbital period of $3.9227090^{+0.0000017}_{-0.0000018}$ d, which is roughly $1.5\sigma$ shorter than the recently published period measurement in \citet{ivshina2022}: $3.92271159 \pm 0.00000038$ days. The Spitzer transits of WASP-29b were previously analyzed by \citet{baxter2021}, who obtained 3.6 and 4.5 $\mu$m transit depths of $9500 \pm 100$ and $9300 \pm 100$ ppm, respectively, which agree with our measurements at better than the $1\sigma$ level.

For WASP-80b, the literature values of $a/R_{*}$ span a wide range from $12.0647 \pm 0.0099$ \citep{sedaghati2017} to $12.99 \pm 0.03$ \citep{triaud2013}; our measurement of $12.451^{+0.073}_{-0.071}$ lies intermediate to those two extremes and within $1.6\sigma$ of the values presented in \citet{kirk2018} and \citet{triaud2015}---$12.66^{+0.12}_{-0.11}$ and $12.63^{+0.08}_{-0.13}$, respectively. The orbital inclination $i$ we derived from our fits is somewhat lower than the range of values in the literature, lying roughly $2.1\sigma$ below the inclination reported in \citet{kirk2018}. For the orbital period, our result---$3.0678569 \pm 0.0000016$ days---agrees well with the most precise period value in the literature: $3.06785500 \pm 0.00000036$ days \citep{parviainen2018}. The Spitzer 3.6 and 4.5 $\mu$m transits of WASP-80b were previously analyzed in \citet{triaud2015}, who reported $R_{p}/R_*$ values that are 1.4 and 2.5$\sigma$ smaller than our measurements. This discrepancy is consistent with the significantly smaller inclination value we obtained from our fits.

\begin{figure*}[t!]
\begin{center}
\includegraphics[width=\linewidth]{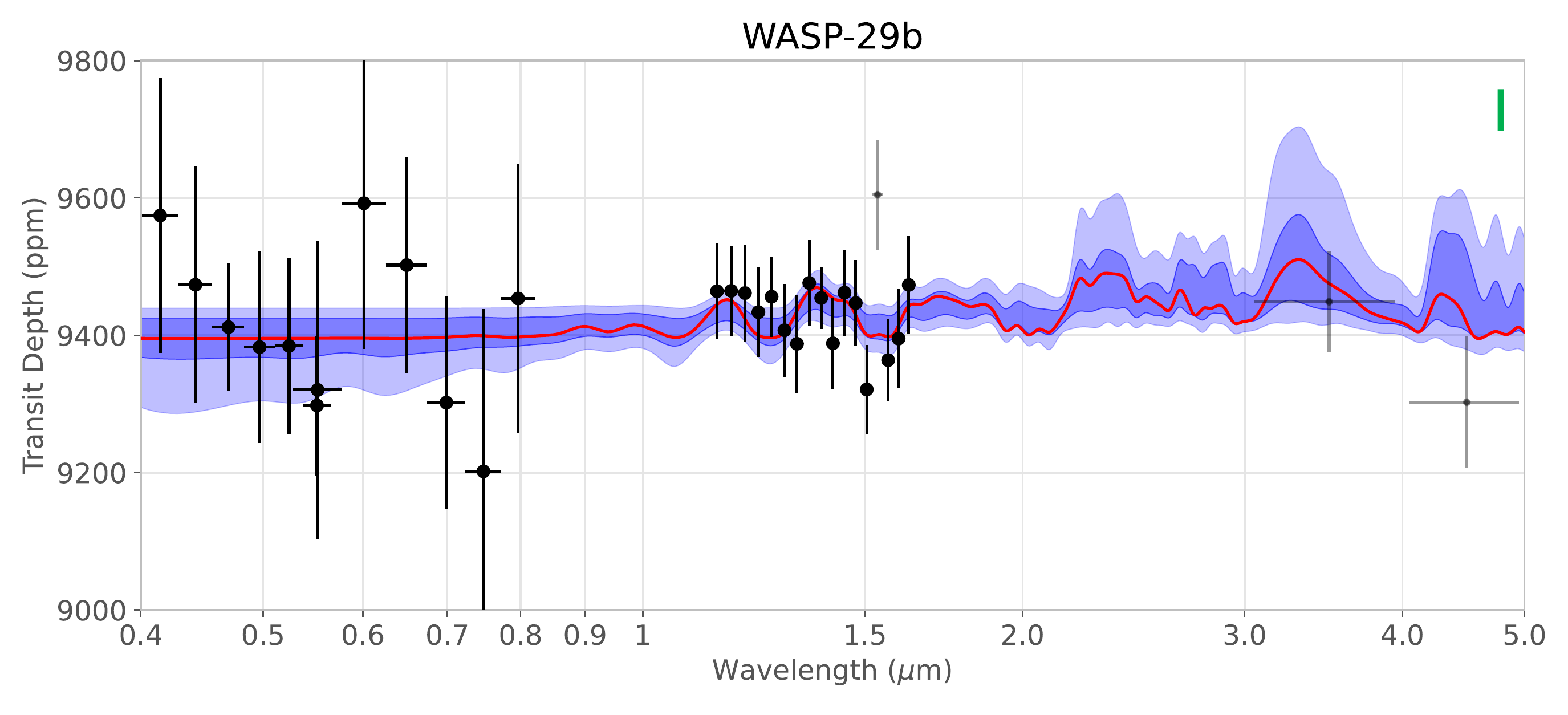}
\end{center}
\caption{Measured transmission spectrum of WASP-29b (black data points) alongside the result of our PLATON atmospheric retrieval. The red curve shows the best-fit atmospheric model, while the medium and light blue shaded regions indicate the 1$\sigma$ and 2$\sigma$ bounds for the atmospheric models, respectively. The mostly flat transmission spectrum is consistent with a cloudy atmosphere without fine-particle aerosols and a wide range of atmospheric metallicities. Our preferred model includes an additive offset for the STIS points relative to the WFC3 points, which has been included in the plotted visible-wavelength points. The low Spitzer 4.5 $\mu$m transit depth is inconsistent with physically reasonable atmospheric models, and both Spitzer points (gray points) were excluded from the retrieval analysis, along with the outlier point at $\sim$1.5 $\mu$m. The green bar denotes the atmospheric scale height corresponding to the best-fit model from our PLATON retrieval (52 ppm).} \label{wasp29retrieval}
\vspace{0cm}
\end{figure*}


\subsection{Transmission Spectra}\label{subsec:spectra}


Given the consistency between the broadband transit depths measured from the two STIS G430L visits of each target, we simply applied a weighted average to the two depths measured in each spectroscopic bin when constructing the final combined transmission spectra. At the blue end of the G430L grating and the red end of the G750L grating, the measured flux from the star drops considerably, resulting in significantly higher uncertainties in the transit depths and larger scatter in the spectrum. In both the table and the figures, we only present values for wavelength bins in which the transit depth $(R_{p}/R_{*})^2$ was measured to a precision of $<$250 ppm in the case of WASP-29b and $<$400 ppm in the case of WASP-80b. For the WFC3 G141 bandpass, we do not include the transit depths for the bluest and the two reddest 30 nm wavelength bins (1.10--1.13, 1.64--1.67, and 1.67--1.70 $\mu$m), because the corresponding spectroscopic light curves have significantly poorer signal-to-noise ratios relative to the others.

\begin{figure*}[t!]
\begin{center}
\includegraphics[width=\linewidth]{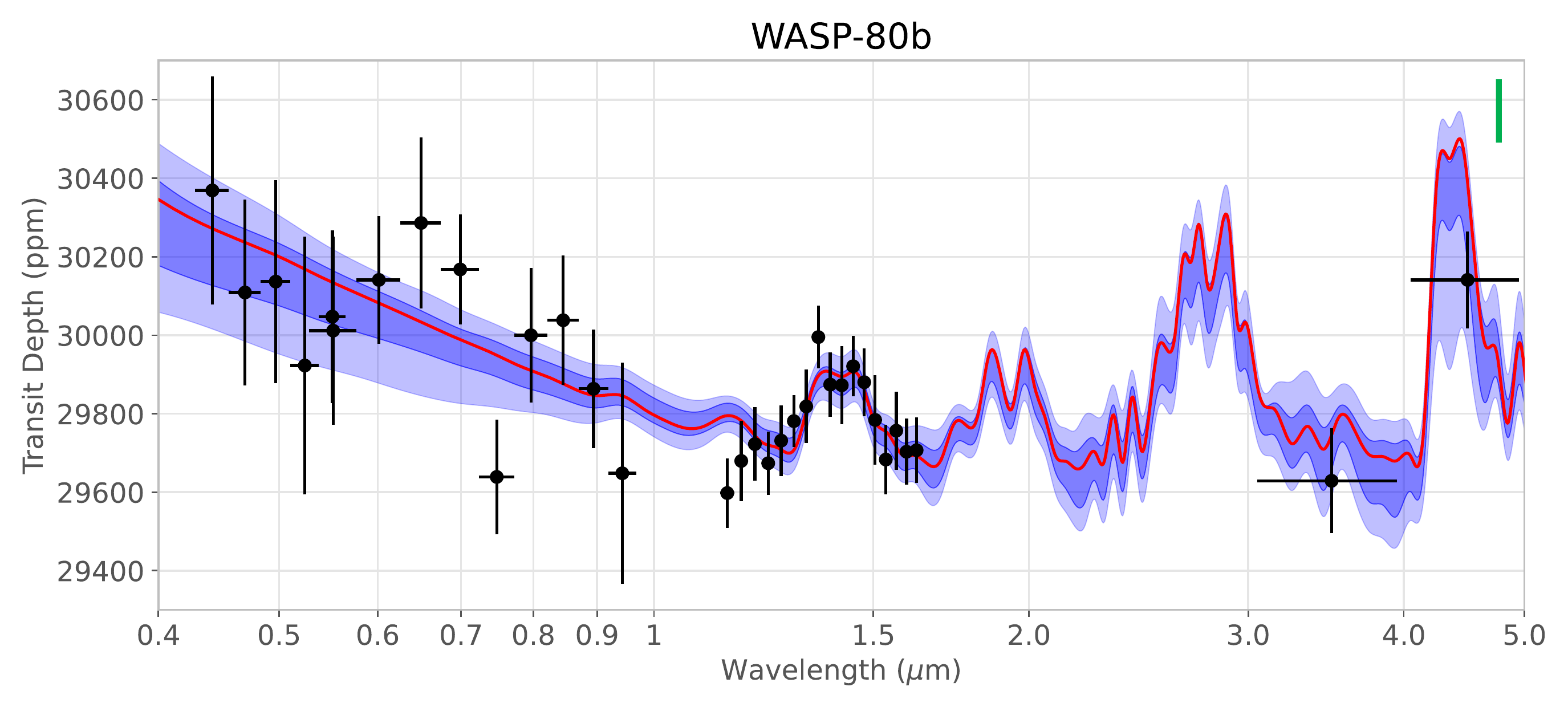}
\end{center}
\caption{Same as Figure~\ref{wasp29retrieval}, but for WASP-80b. In the optical portion of the spectrum, there is strong evidence for a hazy atmosphere with a scattering slope consistent with Rayleigh scattering. The prominent absorption feature in the WFC3 bandpass and the relatively high Spitzer 4.5 $\mu$m transit depth are indicative of H$_2$O and CO$_2$ in the photosphere, respectively. We find that the atmosphere has a near-solar C/O ratio and a cloud deck extending up to the 1 mbar pressure level. The atmospheric scale height (green) from the best-fit PLATON model is 155 ppm.} \label{wasp80retrieval}
\vspace{0cm}
\end{figure*}

The measured planet--star radius ratios from the spectroscopic light-curve analyses are listed in Table~\ref{tab:specfit}. The full 0.4--5.0 $\mu$m transmission spectra of WASP-29b and WASP-80b are shown in Figures~\ref{wasp29retrieval} and \ref{wasp80retrieval}. A compilation of the spectroscopic light curves and corresponding best-fit transit models is provided in the Appendix.

In the STIS G750L bandpass, the transit depths obtained for the narrow wavelength bins centered around the main Na and K absorption regions are broadly consistent with the depths measured in the wider 100-pixel spectroscopic bins spanning those regions, suggesting no significant alkali absorption feature in the transmission spectra of either planet. These narrowband transit depths are not included in the transmission spectrum plots. However, we note that the poor signal-to-noise ratio of these transit-depth measurements makes it difficult to reach definitive conclusions about the presence or absence of alkali absorptions in the transmission spectra. In particular, the relatively deep 577--626 nm transit of WASP-29b may indicate some amount of Na absorption, a possibility that we explore in more detail in Section~\ref{sec:retrieval}.

The reddest wavelength bin in the STIS G430L grating (538--565 nm) and the bluest wavelength bin in the STIS G750L grating (528--577 nm) overlap. For both planets, the transit depths measured from those two spectroscopic light curves are consistent to well within $1\sigma$. This indicates that there is no evidence for any significant common-mode offsets between the G430L and G750L visits.

Looking at Figure~\ref{wasp29retrieval}, we see that the transmission spectrum of WASP-29b is largely consistent with a flat line, with no detected water absorption feature in the WFC3 G141 bandpass and no evidence for a significant Rayleigh scattering slope in the optical. The shape of the optical transmission spectrum is in good agreement with the previously published ground-based 515--720 nm spectrum from \citet{gibson2013}, which is consistent with a flat line and likewise does not contain any significant alkali absorption features. \citet{tsiaras2018} published an independent analysis of the WFC3 G141 transmission spectrum as part of their ensemble study. They also found a featureless 1.1--1.7 $\mu$m spectrum, albeit with a slightly more pronounced negative slope across the bandpass than in our spectrum. Moreover, near-infrared transit depths are systematically larger (by roughly 50--100 ppm), owing to the smaller inclination value that they used as a fixed parameter in their fits ($88\overset{\circ}{.}8$; \citealt{hellier2010}).


In contrast to WASP-29b, we find a robust 1.4 $\mu$m water absorption feature for WASP-80b. As illustrated in Figure~\ref{wasp80retrieval}, the optical part of the transmission spectrum shows significant scatter, while being broadly consistent with a negative spectral slope that suggests Rayleigh scattering by fine-particle aerosols. Several ground-based transmission spectra have been obtained for WASP-80b spanning the visible wavelength range. Focusing on those with spectral resolutions greater than $\lambda/\Delta\lambda = 10$, we find a range of spectral slopes. The 500--900 nm spectrum published by \citet{parviainen2018} is consistent with a flat line, while the 500--900 nm spectrum measured by \citet{kirk2018} shows a negative spectral slope consistent with our spectrum. The 750--1000 nm spectrum published by \citet{sedaghati2017} has a very steep negative slope, possibly caused by uncorrected residual systematics in their VLT/FORS light curves. The WFC3 G141 transmission spectrum published by \citet{tsiaras2018} displays a prominent water absorption feature at 1.4 $\mu$m, consistent in shape with our results. Meanwhile, our measured near-infrared transit depths are roughly 200 ppm larger on average, due to the smaller inclination we obtained from our broadband light-curve fit.

\begin{deluxetable}{lcccc}[t!]
\tablewidth{0pc}
\renewcommand{\arraystretch}{0.9}
\tabletypesize{\small}
\tablecaption{
    Secondary Eclipse Fit Results
    \label{tab:eclipse}
}
\tablehead{\multicolumn{1}{c}{Eclipse} & &
    \multicolumn{1}{c}{Depth (\%)} & &
    \multicolumn{1}{c}{Phase}   
}
\startdata
				 WASP-29b & & & &\\
				3.6 $\mu$m & & &  & \\
				\quad Eclipse 1 & & $0.009 \pm 0.021$ & &$\equiv0.5$\tablenotemark{\scriptsize a} \\
				\quad Eclipse 2 & &$0.011 \pm 0.009$&& $\equiv0.5$\\
			  	\quad Global\tablenotemark{\scriptsize b} & &$0.015 \pm 0.008$ & &$0.5034^{+0.0012}_{-0.0016}$ \\
				4.5 $\mu$m && & &\\
				\quad Eclipse 1 && $0.014 \pm 0.012$ && $\equiv0.5$ \\
				\quad Eclipse 2 && $0.040 \pm 0.009$ && $\equiv0.5$ \\
				 \quad Global\tablenotemark{\scriptsize b} && $0.033 \pm 0.006$ & &$0.5034^{+0.0012}_{-0.0016}$  \\
				\hline
				WASP-80b & & & &\\
				3.6 $\mu$m & & &  & \\
				\quad Eclipse 1 & & $0.025^{+0.012}_{-0.013}$ & &$\equiv0.5$ \\
				\quad Eclipse 2 & &$0.042 \pm 0.011$&& $\equiv0.5$\\
				\quad Eclipse 3 & & $0.041^{+0.010}_{-0.009}$ & &$\equiv0.5$ \\
				\quad Eclipse 4 & &$0.019 \pm 0.008$&& $\equiv0.5$\\
			  	\quad Global\tablenotemark{\scriptsize b} & &$0.028 \pm 0.005$ & &$0.49999 \pm 0.00028$ \\
				4.5 $\mu$m && & &\\
				\quad Eclipse 1 && $0.083 \pm 0.009$ && $\equiv0.5$ \\
				\quad Eclipse 2 && $0.101 \pm 0.009$ && $\equiv0.5$ \\
				 \quad Global\tablenotemark{\scriptsize b} && $0.093^{+0.005}_{-0.006}$ & &$0.49999 \pm 0.00028$ \\
\enddata
\textbf{Notes.}
\vspace{-0.2cm}\tablenotetext{\textrm{a}}{Eclipse phase fixed to 0.5 for all individual eclipse fits, assuming the orbital ephemeris measured from the respective global transit fits (Table~\ref{tab:globalfit}). The reported eclipse phases from the global analyses are corrected for the light travel time across the system.}
\vspace{-0.2cm}\tablenotetext{\textrm{b}}{Computed from a joint fit to all eclipses for each planet.}
\vspace{-1cm}
\end{deluxetable}

\begin{figure}[t!]
\begin{center}
\includegraphics[width=\linewidth]{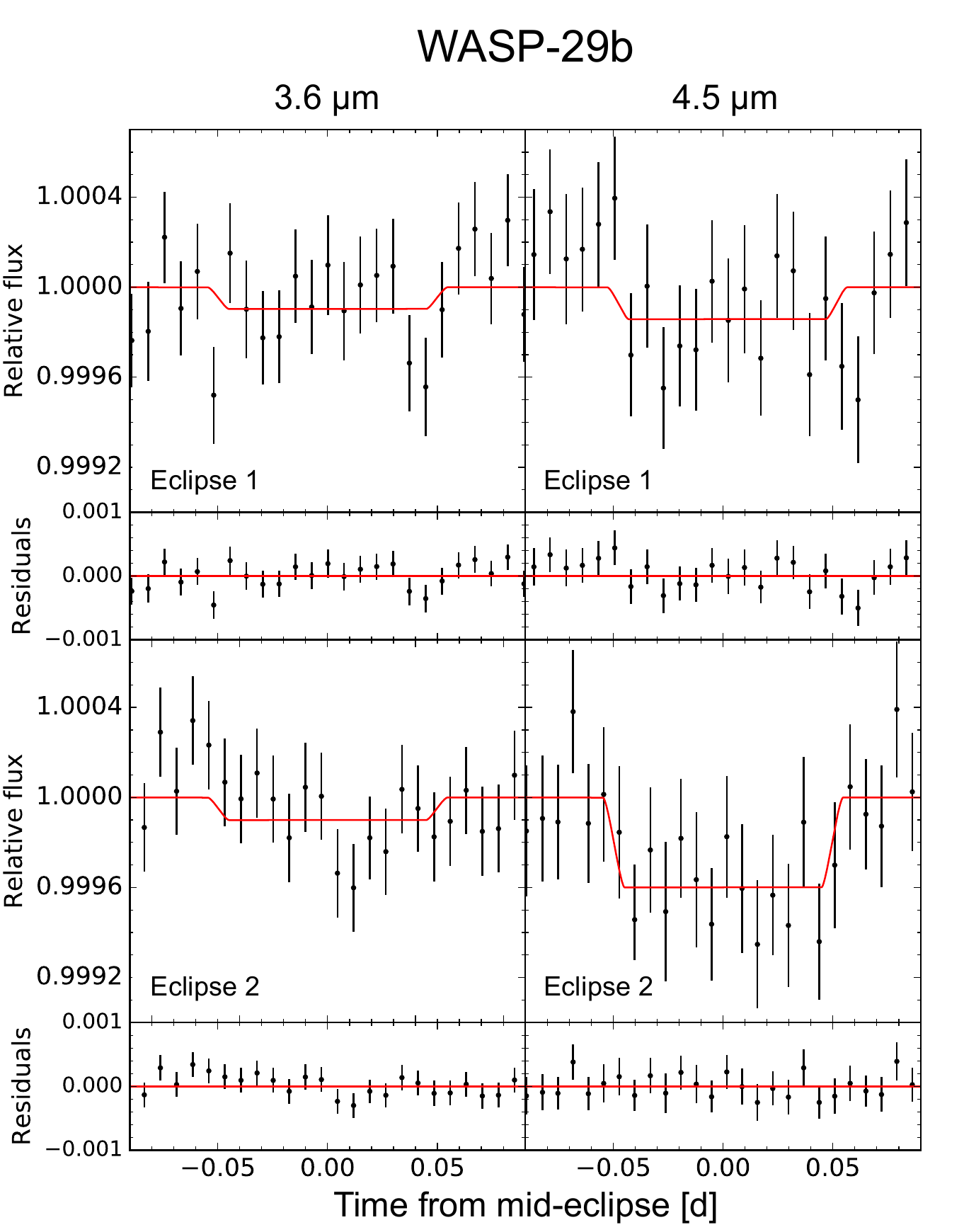}
\end{center}
\caption{The four Spitzer 3.6 and 4.5 $\mu$m secondary eclipse light curves of WASP-29b (black points), binned in intervals of 10 minutes, with the best-fit eclipse model plotted in red. The corresponding residuals are shown beneath earch light curve. Only the second observation at 4.5 $\mu$m shows a robust eclipse detection.} \label{wasp29ecl}
\end{figure}

\begin{figure}[t!]
\begin{center}
\includegraphics[width=\linewidth]{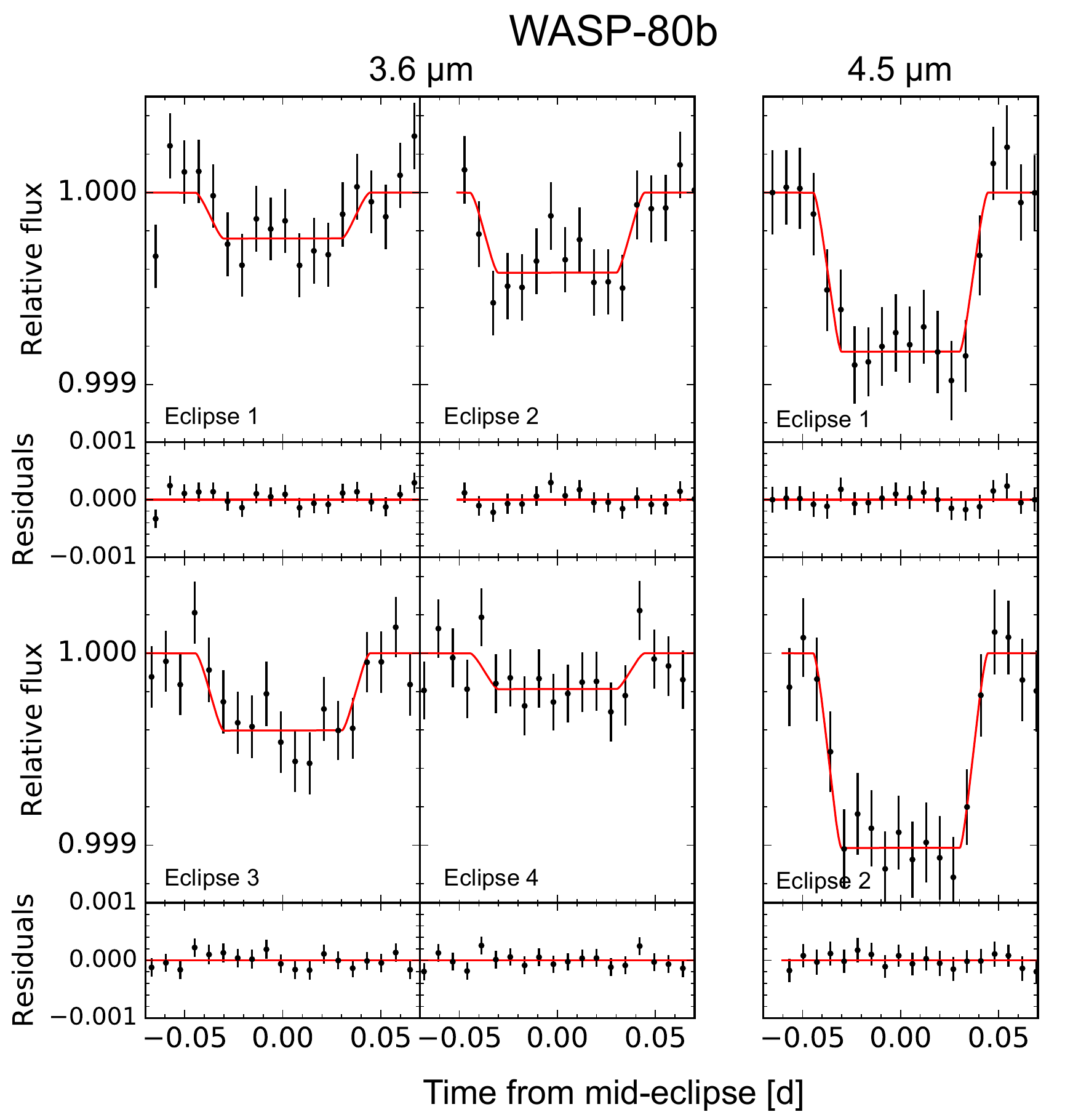}
\end{center}
\caption{Similar to Figure~\ref{wasp29ecl}, but for the six WASP-80b secondary eclipse observations.} \label{wasp80ecl}
\end{figure}

\quad\\
\subsection{Secondary Eclipse Fits}\label{subsec:ecl}

The Spitzer secondary eclipse light curves were fit using the same PLD systematics model as the transits. Due to the low signal-to-noise ratio, we fixed the transit-shape parameters and orbital ephemeris to the best-fit values obtained from our broadband transit light-curve analysis (Table~\ref{tab:globalfit}). When fitting the individual visits to optimize the photometric extraction parameters and binning/trimming, we set the mid-eclipse time to an orbital phase of 0.5. The measured depths for all secondary eclipse light curves are listed in Table~\ref{tab:eclipse}. The binned and systematics-corrected WASP-29b and WASP-80b eclipse light curves are plotted in Figures~\ref{wasp29ecl} and \ref{wasp80ecl}, respectively. Most of the eclipse-depth measurements are marginal, particularly at 3.6 $\mu$m. When comparing the individual eclipse depths in the same bandpass, we find that they are mutually consistent at better than the $2\sigma$ level in all cases.

For each system, we carried out a joint fit to all secondary eclipse observations. In order to probe for nonzero orbital eccentricity, we allowed the mid-eclipse orbital phase to vary freely. All eclipse light curves within a bandpass were assigned a common eclipse depth. The results from our joint fits are presented in Table~\ref{tab:eclipse}. We have corrected the mid-eclipse orbital phases by the light travel time between the superior and inferior conjunctions. For WASP-29b, the best-fit 3.6 and 4.5 $\mu$m eclipse depths are $0.015\% \pm 0.008\%$ and $0.033\% \pm 0.006\%$, respectively, and the mid-eclipse orbital phase is $0.5034^{+0.0012}_{-0.0016}$, roughly $2.1\sigma$ after mid-orbit. For WASP-80b, the eclipse depths are $0.028\% \pm 0.005\%$ and $0.093\%^{+0.005\%}_{-0.006\%}$, respectively; the best-fit eclipse phase is consistent with 0.5 at better than the $1\sigma$ level, indicating that the orbit is consistent with circular. 

\citet{triaud2015} presented an independent analysis of the WASP-80b eclipses obtained as part of their Spitzer program, i.e., Eclipses 1 and 2 in each bandpass, as listed in Table~\ref{spitzerphotometry}. From a global analysis of the Spitzer transit and secondary eclipse light curves, along with several other ground-based transit observations, they obtained best-fit eclipse depths of $0.0455\% \pm 0.0100\%$ and $0.0944\%^{+0.0064\%}_{-0.0065\%}$ at 3.6 and 4.5 $\mu$m, respectively. Their 4.5 $\mu$m depth agrees with our value to well within $1\sigma$, while their 3.6 $\mu$m measurement is larger than ours by roughly $1.6\sigma$. We reran the global eclipse light-curve analysis without the third and fourth 3.6 $\mu$m secondary eclipse light curves and obtained a depth of $0.35\% \pm 0.09\%$, which is consistent with the \citet{triaud2015} value at better than the $1\sigma$ level.

From the global eclipse depths listed in Table~\ref{tab:eclipse}, we derived the corresponding dayside brightness temperatures, following the methodology described in \citet{wong2020}. We computed the stellar flux at 3.6 and 4.5 $\mu$m, integrated across the Spitzer bandpasses, for a grid of PHOENIX stellar models and fit for polynomial functions with respect to stellar brightness temperature, metallicity, and surface gravity. These empirical relationships were used to self-consistently propagate the uncertainties in the measured stellar parameters (Table~\ref{tab:properties}) to our estimates of the planet's dayside brightness temperature. The eclipse depth depends on the star--planet radius ratio as well, and we utilized the most precise value from our broadband light-curve analysis---the transit depth in the WFC3 G141 bandpass. Because the primary contribution to the uncertainty in the brightness temperature is the eclipse-depth uncertainty, choosing any of the other listed $R_{p}/R_{*}$ values in Table~\ref{tab:globalfit} did not significantly affect our temperature estimates.

The 3.6 and 4.5 $\mu$m dayside brightness temperatures for WASP-29b are $830^{+150}_{-440}$ and $940^{+48}_{-52}$ K, respectively. Fitting a single blackbody to both eclipse depths yielded a dayside brightness temperature of $937^{+45}_{-48}$ K. For WASP-80b, the dayside brightness temperatures in the two Spitzer bandpasses are $791^{+28}_{-30}$ and $871^{+15}_{-16}$ K, respectively. The two eclipse depths are consistent at better than $2\sigma$ with a single blackbody at $851^{+13}_{-14}$ K. For both WASP-29b and WASP-80b, our measured dayside brightness temperatures are in excellent agreement with the equilibrium temperatures listed in Table~\ref{tab:properties} ($970^{+32}_{-31}$ and $825 \pm 19$ K, respectively). The equilibrium temperatures were calculated assuming zero Bond albedo and no heat recirculation to the nightside. Therefore, we conclude that both planets have weak day--night heat transport.

General circulation models of temperate planets with similar equilibrium temperatures to WASP-29b and WASP-80b have generally predicted more efficient global heat transport and correspondingly reduced day--night temperature contrasts when compared to their hotter counterparts \citep[e.g.,][]{showman2015}. Therefore, at first glance, our inference of relatively weak day--night heat transport for these two planets stands in contradiction to the prevailing trend. However, metallicity has been shown to also affect the global energy budget of planetary atmospheres, with some studies showing that higher metallicities inhibit day--night heat recirculation \citep[e.g.,][]{zhang2017}. The enhanced metallicities that we retrieve from the transmission spectra (Section~\ref{sec:retrieval}) may indicate that the metal enrichment is weakening longitudinal heat transport. 

More broadly, there is a marked dearth of relatively cool exoplanets that have been examined in detail with sophisticated atmospheric modeling \citep[e.g.,][]{zhang2020rev}. This has primarily been caused by the lack of high-quality emission data for these targets. The temperature distribution of warm gas giants such as WASP-29b and WASP-80b has hitherto only been constrained with broadband Spitzer observations. The James Webb Space Telescope (JWST) will enable intensive spectroscopic study of these cooler objects and definitively probe fundamental trends in atmospheric heat transport in this temperature regime.

\section{Atmospheric Retrievals}\label{sec:retrieval}

To explore the atmospheric properties of WASP-29b and WASP-80b, we utilized the PLanetary Atmospheric Transmission for Observer Noobs (PLATON) retrieval suite \citep{zhang2019,zhang2020}. PLATON is an open-source Python-based retrieval code that combines robust Bayesian inference algorithms with fast atmospheric forward modeling, which is largely based on the Exo-Transmit framework \citep{kempton2017}. For given values of atmospheric metallicity $\log (Z/Z_{\odot})$ and C/O ratio, PLATON uses GGchem \citep{woitke2018} to calculate the relative abundances of 34 potentially relevant chemical species (including H$_2$O, CO, CO$_2$, CH$_4$, NH$_3$, HCN, Na, K, TiO, and VO) at each pressure layer, assuming chemical equilibrium and accounting for the effects of condensation. Depending on the dataset being fit (transmission or emission spectra), the user can retrieve either an isothermal temperature or a parameterized TP profile. Cloud opacity is handled via a gray cloud deck with a cloudtop pressure $P_{\rm cloud}$ that blocks all impinging radiation. In addition, haze scattering can be introduced, with options for either a simple Rayleigh-like scattering slope or a more detailed microphysical Mie scattering calculation. After populating the abundance--temperature--pressure grid, PLATON computes the total opacity at each pressure level, converts the pressure levels to heights assuming hydrostatic equilibrium and the ideal gas law, and finally derives both the transmission and dayside emission spectra using standard one-dimensional radiative transfer routines. 

As discussed in Section~\ref{subsec:activity}, starspots can substantively affect the shape of the measured transmission spectrum. PLATON accounts for unocculted starspots by modeling the stellar surface as a mixture of an unspotted photosphere emitting at the effective temperature $T_{\rm eff}$ and starspots emitting at a lower temperature $T_{\rm spot}$ that cover a fraction $f_{\rm spot}$ of the Earth-facing stellar disk \citep[e.g.,][]{evans2018,zhang2019}. 

In our retrieval analysis of WASP-29b and WASP-80b, we only considered the measured transmission spectra. This choice was motivated by the relatively low signal-to-noise ratio of the broadband Spitzer secondary eclipse depths. Because the TP profile and atmospheric composition across the dayside hemisphere are expected to differ from the corresponding properties along the limb, the emission spectrum cannot be reliably used to provide additional constraints for the transmission spectrum retrieval. Meanwhile, our occultation measurements lack sufficient wavelength coverage, spectral resolution, and precision to adequately constrain the dayside atmosphere on their own.

The current version of PLATON (v5.2.1) includes both the standard MCMC ensemble sampler \texttt{emcee} and the multimodal nested sampler \texttt{dynesty} \citep{speagle2020}. Nested sampling allows for highly efficient exploration of the posterior distributions, particularly those that are non-Gaussian, and automatically determines when convergence is achieved. The nested sampling routine also provides the Bayesian evidence of the model, $\mathcal{Z}$, which quantifies both the goodness of fit and model complexity to prevent overfitting (i.e., excessive model structure; see, for example, \citealt{trotta2017}). In our retrieval runs, we used the default nested sampler routine in PLATON (with multi-ellipsoidal bounds and random-walk sampling) and considered the Bayesian evidence, along with physical arguments, when selecting the optimal model.

For both planets, we experimented with a range of models, starting from the fiducial case of a haze-free isothermal atmosphere with metallicity $\log(Z/Z_{\odot})$, C/O ratio, and cloudtop pressure $\log P_{\rm cloud}$ as the primary free parameters to be retrieved. We then incrementally added or removed model complexity and compared the resultant Bayesian evidence and other fit-quality metrics to select the optimal model. As a sanity check, we also carried out retrievals on subsets of the data (e.g., WFC3 and STIS only) to examine the importance of each dataset in constraining the atmospheric properties and probe for anomalous data points that may skew the retrieved parameters. Furthermore, we considered systematic deviations between the transit depths from the various instruments by introducing additive instrumental offsets (defined relative to the WFC3 values). Such offsets can arise from our use of common-mode systematics correction when detrending the spectroscopic light curves, as well as uncertainties in the transit-shape parameters (see discussion in \citealt{sheppard2021}). The offset parameters were constrained by Gaussian priors, with the widths set to the uncertainties in the corresponding broadband transit depths (Table~\ref{tab:globalfit}). 

\begin{deluxetable}{lcc}[t!]
\tablewidth{0pc}
\setlength{\tabcolsep}{5pt}
\renewcommand{\arraystretch}{0.9}
\tabletypesize{\footnotesize}
\tablecaption{
    Parameters and Priors Used in PLATON Atmospheric Retrievals
    \label{tab:priors}
}
\tablehead{ \vspace{-0.2cm}\\ \multicolumn{1}{c}{Parameter} & 
    \multicolumn{1}{c}{Symbol} &
    \multicolumn{1}{c}{Prior\tablenotemark{\scriptsize a} }   
}
\startdata
Planet radius ($R_{\rm Jup}$)& $R_{p}$ & ${\mathcal U}(0,1.5)$\\
Planet mass ($M_{\rm Jup}$) & $M_{p}$ & ${\mathcal N}(0.244,0.020)$\\
 & & ${\mathcal N}(0.538,0.036)$ \\
Stellar radius ($R_{\odot}$) & $R_{*}$ & ${\mathcal N}(0.808,0.044)$\\
 & & ${\mathcal N}(0.586,0.018)$ \\
Stellar temperature (K) & $T_{\rm eff}$ & ${\mathcal N}(4800,150)$ \\
 & & ${\mathcal N}(4145,100)$ \\
Limb temperature (K) & $T$ & ${\mathcal U}(300,1100)$\\
Atmospheric metallicity  & $\log (Z/Z_{\odot})$ & ${\mathcal U}(-1,3)$ \\
Carbon--oxygen ratio & C/O & ${\mathcal U}(0.2,2.0)$ \\
Sodium abundance & $\log {\rm Na}$ & ${\mathcal U}(-12,-3)$ \\
Cloudtop pressure (Pa) & $\log P_{\rm cloud}$ & ${\mathcal U}(-1,6)$ \\
Scattering factor & $\log \alpha_{s}$ & ${\mathcal U}(-5,5)$\\
Scattering slope & $\gamma_{s}$ & ${\mathcal U}(0,10)$\\
STIS offset (ppm) & $\xi$ & ${\mathcal N}(0,100)$ \\
 & & ${\mathcal N}(0,150)$ \\
Mie particle size (m) & $\log r_{\rm part}$ & ${\mathcal U}(-9,-3)$ \\
Mie number density (m$^{-3}$) & $\log n$ & ${\mathcal U}(0,10)$ \\
Fractional scale height  & $H_{\rm cloud}/H_{\rm gas}$ & ${\mathcal U}(0.5,7)$\\
Starspot coverage fraction\tablenotemark{\scriptsize b} & $f_{\rm spot}$ & ${\mathcal U}(0,1)$ \\
Error multiple & $\beta$ & ${\mathcal U}(0.5,5)$\\
\enddata
\textbf{Notes.}
\vspace{-0.2cm}\tablenotetext{\textrm{a}}{The symbols ${\mathcal N}$ and ${\mathcal U}$ denote normal (log-normal in the case of logarithmic variables) and uniform distributions, respectively. For normal priors, the median and $1\sigma$ values are provided; for uniform priors, the lower and upper bounds are given. Priors for the WASP-29b and WASP-80b retrieval runs are listed separately whenever they differ.}
\vspace{-0.2cm}\tablenotetext{\textrm{b}}{Retrieval runs including starspots were only carried out for WASP-80b. The spot temperature was fixed to $T_{\rm spot} = 3350$ K.}
\vspace{-0.5cm}
\end{deluxetable}

The relatively high WASP-29b transit depth at 577--626 nm motivated us to alter the PLATON code to allow for a freely varying sodium abundance ($\log {\rm Na}$). In those instances, the sodium abundance was uniformly set at all pressure levels, while the other atmospheric components remained at their equilibrium (pressure-dependent) abundances.

Table~\ref{tab:priors} lists all of the parameters that were allowed to vary in our suite of PLATON retrieval runs, along with the applied priors. For the planet mass, stellar radius, and stellar effective temperature, the Gaussian (i.e., normal) priors were defined based on the literature values listed in Table~\ref{tab:properties}. For the planetary radius, we fixed the reference pressure level to 1 bar and did not apply a Gaussian prior based on the literature measurements. For most of the other parameters, the prior definitions were identical to the default settings in PLATON. For Rayleigh scattering, the absorption coefficient is defined as $\alpha_s\lambda^{-\gamma_s}$. The prefactor $\alpha_s$ quantifies the strength of the haze (i.e., fine-particle aerosols) scattering relative to pure Rayleigh scattering at the reference wavelength of 1 $\mu$m; the slope of the scattering opacity $\gamma_{s}$ can vary from 0 to 10, with the default value 4 corresponding to the standard Rayleigh-scattering case. Meanwhile, for Mie scattering, we considered Titan tholins \citep{khare1984} and hydrocarbon soots \citep{morley2015,lavvas2017,gao2020}. PLATON is not designed to allow for wavelength-dependent user-defined refractive indices, so we instead set the real and imaginary refractive indices to the averages of the literature values at wavelengths shorter than 1 $\mu$m (where the effects of haze opacity on the transmission spectrum shape are most prominent). We allowed the mean particle size $\log r_{\rm part}$, number density $\log n$, and fractional scale height $H_{\rm cloud}/H_{\rm gas}$ to vary. The distribution of particle sizes is log-normal, with the geometric standard deviation set to 0.5 by default. Finally, we included an error multiple $\beta$ that was applied uniformly across the entire transmission spectrum in order to account for over- or underestimation of the transit-depth uncertainties.

\begin{deluxetable}{lccccl}[t!]
\tablewidth{0pc}
\setlength{\tabcolsep}{3pt}
\renewcommand{\arraystretch}{0.9}
\tabletypesize{\scriptsize}
\tablecaption{
    Model Comparison for PLATON Atmospheric Retrievals \label{tab:retrievalcomp}
}
\tablehead{ \\ Model &  $\ln(\mathcal{Z}_i$)\tablenotemark{\scriptsize a} & $\ln(\mathcal{L}_i$)\tablenotemark{\scriptsize a} &  $\chi_r^2$\tablenotemark{\scriptsize a} & DOF\tablenotemark{\scriptsize a} &
$B_i$\tablenotemark{\scriptsize a}
}
\startdata
\multicolumn{2}{l}{WASP-29b (no Spitzer points)} \\
fiducial + offset      & $219.47 \pm 0.18$   &      234.11 & 0.56 & 18 &  --- \\
no offset              & $218.36 \pm 0.19$  &       232.56 & 0.57 & 19 & 3.03 ($2.1\sigma$) \\
with hazes             & $219.36 \pm 0.20$  &      235.77 & 0.58 & 17 & 1.12 \\
with hazes \\
\, + free scattering slope & $219.40 \pm 0.21$  &  236.56 & 0.56 & 16 & 1.07 \\
with hazes + free Na  & $219.44 \pm 0.19$ & 236.17 & 0.60 & 16 & 1.03 \\
no clouds              & $218.61 \pm 0.19$ &              233.86 & 0.53 & 19 &  2.36 ($1.9\sigma$) \\
\hline
WASP-80b\\
fiducial + hazes      & $240.71 \pm 0.20$   &              255.99 & 0.91 & 23 &  --- \\
with offset & $240.69 \pm 0.23$           &              256.02  & 0.88 & 22 & 1.02 \\
no ${\rm H_{2}O}$   & $233.55 \pm 0.17$   &              246.01 & 1.71 & 23 & 1290 ($4.2\sigma$) \\
no ${\rm CO}$     & $239.99 \pm 0.21$   &              255.65 & 0.93 & 23 &  2.05 \\
no ${\rm CO_2}$   & $236.48 \pm 0.20$   &              251.90 & 1.17 & 23 & 68.7 ($3.4\sigma$) \\
no ${\rm CH_4}$   & $240.59 \pm 0.20$   &              255.51 & 0.93 & 23 & 1.13 \\
no clouds & $240.48 \pm 0.19$   &              255.26 & 0.91 & 24 & 1.26 \\
no hazes  & $230.21 \pm 0.17$   &             241.62 & 2.09 & 24 &  36300 ($5.0\sigma$) \\
free scattering slope & $240.61 \pm 0.23$ &               255.70 & 0.97 & 22 & 1.11 \\
Mie scattering (tholins) & $235.76 \pm 0.25$ &                      254.60 & 1.08 & 21 & \dots\tablenotemark{\scriptsize b} \\   
Mie scattering (soots) & $234.30 \pm 0.28$ &                      253.93 & 1.11 & 21 & \dots\tablenotemark{\scriptsize b} \\   
no hazes, with starspots & $237.50 \pm 0.22$ &            255.57 & 0.95 & 23 & \dots\tablenotemark{\scriptsize b} \\
\enddata
\textbf{Note.}
\vspace{-0.2cm}\tablenotetext{\textrm{a}}{$\ln (\mathcal{Z}_{i})$: Bayesian evidence and uncertainty; $\ln(\mathcal{L}_i$): maximum log-likelihood; $B_{i}\equiv \mathcal{Z}_{\rm max}/\mathcal{Z}_i$: Bayes factor relative to the best-performing model, which is listed first for each planet; $\chi_r^2$: reduced chi-squared of the best-fit model, computed without the error multiple parameter $\beta$; DOF: degrees of freedom.}
\vspace{-0.2cm}\tablenotetext{\textrm{b}}{Due to the distinct parameterizations and correspondingly different prior volumes of these models, the relative Bayesian evidence and Bayes factor cannot be readily compared.}
\vspace{-1cm}
\end{deluxetable}

\subsection{WASP-29b}\label{subsec:wasp29retrieval}

The mostly flat and featureless transmission spectrum of WASP-29b indicates the presence of clouds and/or very high atmospheric metallicity that attenuate the absorption features that would otherwise be detectable, particularly in the WFC3 G141 bandpass. Our retrieval runs invariably returned cloudtop pressures in the range 0.1--1 mbar. A few complications arose during our retrieval analysis. The first is the low Spitzer 4.5 $\mu$m transit depth, which pushed the limb temperature posterior against the lower edge of the temperature range allowed by PLATON (300 K). Given that WASP-29b has an equilibrium temperature of roughly 970 K (Table~\ref{tab:properties}), a limb temperature at or below 300 K is physically implausible. The second complication is the low average transit depth across the STIS bandpasses. In particular, the average level of the visible transmission spectrum is lower than the average depth in the WFC3 G141 bandpass. As demonstrated by our comparisons of retrieval runs that included or excluded the STIS transmission spectrum (see below), the relatively low optical transit depths push the limb temperature down to very low values, while simultaneously forcing a mostly cloud-free atmosphere and extremely high metallicity. 

To address the issue of the STIS transit depths, we included an offset (with a prior standard deviation of 100 ppm) and ran retrievals on the transmission spectrum without the Spitzer points. The resultant limb temperatures were more reasonable. However, when we reintroduced the Spitzer data points into the retrieval and allowed for offsets between the Spitzer depths and the WFC3 points, we still found a strong preference for extremely low temperatures. Therefore, we decided to exclude both Spitzer transit depths from the retrieval analysis presented here. We also removed the high outlier point at 1.52--1.55 $\mu$m.

\begin{deluxetable}{lcc}[t]
\setlength{\tabcolsep}{10pt}
\tablewidth{0pc}
\renewcommand{\arraystretch}{0.9}
\tabletypesize{\small}
\tablecaption{
    PLATON Atmospheric Retrieval Results
    \label{tab:platonres}
}
\tablehead{ & \underline{WASP-29b} & \underline{WASP-80b} \\
\multicolumn{1}{c}{Parameter} &
    \multicolumn{1}{c}{Value} &
    \multicolumn{1}{c}{Value}   
}
\startdata
$R_p$ ($R_{\rm Jup}$) & $0.747 \pm 0.036$ & $0.964 \pm 0.025$ \\
$T$ (K) & $530^{+260}_{-140}$  & $930^{+120}_{-110}$ \\
$\log (Z/Z_{\odot})$ & $1.1 \pm 1.4$ & $1.97^{+0.26}_{-0.51}$ \\
C/O & $1.20^{+0.51}_{-0.60}$ & $0.50^{+0.14}_{-0.17}$ \\
$\log P_{\rm cloud}$ (Pa) & $1.8 \pm 1.1$ & $2.4^{+2.4}_{-0.9}$ \\
$\log \alpha_{s}$ &  \dots & $3.62 \pm 0.48$ \\
$\xi$ (ppm) & $78 \pm 37$ & \dots \\
$\beta$ & $0.67 \pm 0.11$ & $0.89 \pm 0.13$\\
\enddata
\vspace{-0.8cm}
\end{deluxetable}

Table~\ref{tab:retrievalcomp} compiles the Bayesian evidence, maximum likelihood value, reduced $\chi^2$, and degrees of freedom of some of our retrievals. Due to the large uncertainties in the optical portion of the transmission spectrum, the reduced $\chi^2$ values are much lower than unity, indicating that the level of photometric uncertainty is preventing the retrievals from distinguishing among different atmospheric models. The extent to which a given model is disfavored relative to the preferred model (i.e., the run with the highest Bayesian evidence) is quantified by the Bayes factor $B$. The fiducial model with a STIS offset performed the best. Removing the STIS offset yielded a lower Bayesian evidence and lower likelihood and is disfavored with a Bayes factor of 3.03, which approximately corresponds to a $2.1\sigma$ significance level. We experimented with adding fine-particle hazes, but the poorly constrained shape of the transmission spectrum at optical wavelengths means that a wide range of scattering factors and slopes is allowed, with no significant improvement to the likelihood. Meanwhile, a cloud-free atmosphere is disfavored at the $1.9\sigma$ level.

When allowing for both free sodium abundance and hazes, we retrieved a broad range of $\log\,{\rm Na}$ values as high as $-$4. For comparison, the equilibrium sodium abundance at pressures below 1 mbar in the best-fit fiducial model is $\log\,{\rm Na}<-8$. While the fit quality is somewhat improved, the increase in model complexity is penalized by the Bayesian evidence. Notably, even at the upper end of the $\log\,{\rm Na}$ posterior, the model spectrum does not show a very prominent sodium absorption feature. This can be attributed to the necessity for high metallicity and/or significant cloud cover in order to match the null water vapor feature at 1.4 $\mu$m, which works to attenuate the sodium absorption as well. One would have to simultaneously invoke relatively low atmospheric metallicity, depleted water abundance, and significantly enhanced sodium abundance to yield both a strong sodium absorption feature and a flat near-infrared transmission spectrum. Given the poor precision of the measured optical transit depths and the general observation that the overall HST STIS transmission spectrum is consistent with a flat line to within the error bars, we did not pursue the issue of sodium abundance further. More intensive observations (e.g., using large ground-based facilities) are needed in order to resolve the transmission spectrum in the vicinity of the sodium absorption feature.

In Figure~\ref{wasp29retrieval}, we show the results of our preferred retrieval run. The $2\sigma$ confidence region is consistent with completely flat transmission spectra. The retrieved parameter values are listed in Table~\ref{tab:platonres}, while the corner plot of relevant parameters can be found in the Appendix (Figure~\ref{wasp29corner}). The retrieval returned a positive offset of $78 \pm 37$ ppm for the STIS transit depths. The planetary radius estimate of $0.747 \pm 0.036$ $R_{\rm Jup}$ is consistent with the literature value ($0.776 \pm 0.043$ $R_{\rm Jup}$; \citealt{gibson2013}). Due to the low precision of the transit depths at visible wavelengths and the lack of robust spectroscopic features, the limb temperature is poorly constrained, with the $1\sigma$ region spanning 390--790 K. Likewise owing to the lack of discernible absorption features amid significant cloud opacity, the full range of C/O ratios is allowed, including near-solar values and ${\rm C/O} > 1$.

Overall, our retrievals indicate cloudy limbs ($P_{\rm cloud}$ from $<$0.01 mbar to $\sim$1 mbar) and unconstrained atmospheric metallicity ($\log (Z/Z_{\odot}) = 1.1 \pm 1.4$). Inspecting the corner plot, we find a slight hint of the classic degeneracy between cloudtop pressure and atmospheric metallicity: the $3\sigma$ confidence region includes scenarios with cloud-free limbs and very high metallicity ($>$$500\times$ solar). Interior modeling of gas giants has revealed a broad range of possible atmospheric metallicities, contingent upon factors such as the level of internal mixing, with the upper limit set at the planet's bulk metallicity value. The bulk metallicity of WASP-29b has a $2\sigma$ upper limit of $190\times$ solar \citep{thorngren2019}, therefore we consider the cloud-free scenario to be highly unlikely.

\subsection{WASP-80b}\label{subsec:wasp80retrieval}

The transmission spectrum of WASP-80b stands in stark contrast to the WASP-29b spectrum and contains both a prominent scattering slope in the optical and a pronounced absorption feature around 1.4 $\mu$m. We found that the fiducial model with enhanced Rayleigh scattering was the best-performing model out of the various retrieval runs. Looking at Table~\ref{tab:retrievalcomp}, we can see that the model without fine-particle aerosol opacity is strongly disfavored ($B = 36300$, $5.0\sigma$), while allowing the scattering slope to vary did not improve the likelihood or Bayesian evidence. Likewise, employing the more complicated Mie scattering model yielded a slightly poorer fit to the visible portion of the transmission spectrum, with Titan tholins performing slightly better than soots. Both Mie scattering retrieval runs indicated particle sizes less than 0.1 $\mu$m, in agreement with the observed Rayleigh scattering slope throughout the STIS bandpass. Regardless of the choice of scattering parameterization, other important atmospheric parameters such as atmospheric metallicity, C/O ratio, and cloudtop pressure did not change appreciably, indicating that those properties are not primarily determined by the visible portion of the transmission spectrum. Similarly, the inclusion of a common-mode offset for the STIS transit depths is marginally disfavored by the Bayesian evidence, while yielding an insignificant improvement to the maximum log-likelihood and reduced chi-squared.

The results of the preferred retrieval model are shown in Table~\ref{tab:platonres} and Figure~\ref{wasp80retrieval}; the corner plot is displayed in Figure~\ref{wasp80corner}. The quality of the fit is excellent, and the introduction of offsets between the various instrumental datasets did not yield improved fits. The water absorption feature in the WFC3 bandpass is well-modeled, as is the relatively deep Spitzer 4.5 $\mu$m transit depth, which suggests CO and/or CO$_2$ absorption. To quantify the significance of the molecular detections, we carried out a series of retrievals, each time removing the opacity contribution from one major atmospheric species (selected from among H$_2$O, CO, CO$_2$, and CH$_4$). This process did not affect the equilibrium abundances calculated by PLATON, but instead simply set the corresponding absorption cross section to zero. The Bayesian evidences of these retrieval runs were then compared to the fiducial model with hazes. As shown in Table~\ref{tab:retrievalcomp}, the removal of H$_2$O and CO$_2$ leads to significantly poorer fits, with Bayes factors of 1290 and 68.7, respectively; meanwhile, we did not find a statistically significant spectroscopic contribution from CO or CH$_4$ in the measured transmission spectrum (but see discussion in Section~\ref{subsec:disc}).

The limb temperature of WASP-80b is well-constrained by the transmission spectrum: $930^{+120}_{-110}$ K. This value is consistent with the dayside temperature that we derived from the Spitzer secondary eclipse depths---$851^{+13}_{-14}$ K (Section~\ref{subsec:ecl})---indicating relatively uniform heat distribution across the dayside hemisphere and extending to the limbs. Due to the presence of the H$_2$O absorption feature, the C/O ratio is restricted to values less than 0.9 \citep[e.g.,][]{benneke2015}, but otherwise unconstrained. The atmospheric metallicity is enhanced, with a mean value of roughly $100\times$ solar and a relatively narrow posterior that excludes near-solar metallicities at greater than $3\sigma$ significance. The corner plot shows a significant cloud--metallicity degeneracy. While the retrieval mostly prefers cloudy atmospheres with atmospheric metallicities of 10--$100\times$ solar, cloud-free limbs are still allowed for metallicities of $\sim$100$\times$ solar and higher. This latter observation was also supported by our cloud-free retrieval, which provided a fit quality similar to that of the fiducial model that includes both clouds and hazes and strongly constrained the atmospheric metallicity to values above roughly $80\times$ solar.

The retrieved $1\sigma$ confidence region of WASP-80b's atmospheric metallicity (30--170$\times$ solar) is notably higher than that of any other planet larger than 0.5 $M_{\rm Jup}$. \citet{thorngren2019} reported a bulk metallicity of $0.17 \pm 0.04$, corresponding to a maximum atmospheric metallicity of $30 \pm 7\times$ solar if the planet is fully mixed, with a $2\sigma$ upper limit of $43\times$ solar. This theoretical range is consistent with the lower end of our retrieved metallicities to within $1\sigma$ and suggests that the atmosphere of WASP-80b is relatively well-mixed. This physical argument also disfavors scenarios of cloud-free limbs with atmospheric metallicities above $100\times$ solar.

An important caveat to our retrieved atmospheric metallicity is the critical role played by the large Spitzer 4.5 $\mu$m transit depth: the high CO/CO$_2$ abundance needed to explain this single point strongly drives the retrieved metallicity to enhanced values. To quantify the effect of the Spitzer portion of the transmission spectrum on the retrieved metallicity, we carried out analogous retrievals on the WFC3 and WFC+STIS spectra. From the WFC3-only retrieval, the atmospheric metallicity is much less constrained, displaying a strong degeneracy with the cloudtop pressure; the $1\sigma$ confidence region of atmospheric metallicity is 3--$150\times$ solar, with a median of $20\times$ solar. Meanwhile, the WFC3+STIS spectrum yields a somewhat narrower range of atmospheric metallicities (10--$130\times$ solar), with a higher median of $40\times$ solar. It is apparent that, without the Spitzer points, the transmission spectrum of WASP-80b becomes broadly consistent with a wide range of atmospheric metallicities, from near-solar to several hundreds of times solar, while being consistent with the tighter constraints from the full dataset retrieval. The results from our Spitzer-free retrievals, along with the theoretical arguments detailed in the previous paragraph and the problematic nature of the 4.5 $\mu$m transit depth in the case of WASP-29b, suggest that the 4.5 $\mu$m transit depth of WASP-80b may be overestimated, biasing our retrieved metallicities to unphysically high values. Future JWST observations (see Section~\ref{subsec:disc}) are needed to obtain an accurate and detailed picture of the transmission spectrum in the 3--5 $\mu$m region and definitively constrain the atmospheric metallicity, cloudtop pressure, and CO/CO$_2$ abundances.

Our photometric monitoring of WASP-80 (Section~\ref{subsec:activity}) revealed discernible stellar variability that was synchronous with the star's rotational period, raising the possibility of starspots on the stellar surface. The appreciable activity of WASP-80 is also corroborated by the X-ray emission observations and the detection of a flare (Section~\ref{subsec:xray}). As unocculted starspots can alter the shape of the transmission spectrum in ways that mimic a Rayleigh scattering slope, we experimented with fitting for haze-free atmospheres while including starspots. The starspot temperature was fixed to the predicted value from the empirical linear scaling law in \citet{rackham2019}: $T_{\rm spot} = 3350$ K. We retrieved $f_{\rm spot} = 0.033 \pm 0.005$, which is consistent with the range of modeled spot-covering fractions for K6V stars in the Kepler sample ($0.014^{+0.027}_{-0.007}$; \citealt{rackham2019}). As shown by the log-likelihood values in Table~\ref{tab:retrievalcomp}, unocculted spots provide a comparably good fit to the steep optical slope in the transmission spectrum in the absence of hazes. We note that, due to the distinct parameterization and correspondingly different prior volume of the hazeless unocculted spot model, the Bayesian evidence cannot be straightforwardly compared with that of the fiducial model with hazes. An important observation is that the spot-covering fraction must be precisely adjusted in order to simultaneously create the apparent Rayleigh scattering slope and produce transmission features that mimic those with the expected atmospheric temperature. Nevertheless, given the self-consistency between the measured photometric variability, retrieved spot-covering fraction, and the corresponding impact on the optical transmission spectrum, it is possible that contamination from unocculted starspots may be at least partially responsible for the observed spectral slope.

\subsection{Discussion}\label{subsec:disc}

Looking at the results of our atmospheric retrievals, we find that the featureless transmission spectrum of WASP-29b hinders detailed quantitative comparisons between the two planets. For example, while the atmospheric metallicity of WASP-80b is largely constrained to highly enriched values, the WASP-29b transmission spectrum is consistent with the full range of allowed metallicity values, from subsolar to 1000$\times$ solar. Nevertheless, several salient differences are apparent from the overall shape of the transmission spectra. 

First, the steep optical slope apparent in the WASP-80b transmission spectrum is entirely absent in the WASP-29b dataset. We reiterate that there is some ambiguity in the interpretation of the optical slope for WASP-80b, with unocculted starspots providing a plausible alternate or contributing explanation for the feature. If we assume that unocculted starspots did not significantly affect the measured transmission spectrum, then the optical slope is indicative of Rayleigh scattering from aerosols with particle sizes smaller than $\sim$0.1 $\mu$m. Such haze particles are readily produced by photochemical processes in the upper atmosphere and are expected to be the dominant source of aerosol opacity at temperatures below $\sim$950 K \citep[e.g.,][]{lavvas2017,horst2018,adams2019,gao2020,ohno2020a}. Meanwhile, the flat spectrum of WASP-29b signals the absence of such fine-particle opacity. 

Flat spectra with obscured molecular absorption features are usually attributed to condensate clouds consisting of $>$1 $\mu$m aerosol particles that extend to low pressure levels \citep[e.g.,][]{fortney2005,helling2008,morley2013,charnay2015,barstow2017,gao2018,powell2019}. Similar spectra have been measured for planets spanning a wide range of equilibrium temperatures \citep[e.g.,][]{deming2013,knutson2014,kreidberg2014,singstis,wakeford2017,tsiaras2018,chachan2020,wakeford2020,alam2022}. Silicate clouds (particularly Mg$_2$SiO$_4$) are expected to be the primary condensate species contributing to enhanced opacity at the equilibrium temperature of WASP-29b, and indeed across most of the temperature range from $\sim$900 to $\sim$2000 K \citep[e.g.,][]{gao2020}. However, recent modeling has shown that photochemical hazes composed of hydrocarbon aggregates can grow to larger sizes than the spherical haze particles typically considered in earlier studies, resulting in a weaker wavelength dependence of the optical depth that can mimic the signature of condensate cloud decks \citep[e.g.,][]{adams2019,ohno2020b}. Therefore, we cannot exclude the possibility that the gray opacity observed on WASP-29b is driven, at least partially, by photochemistry. In fact, both WASP-29b and WASP-80b lie in a temperature range where photochemical haze production is expected to contribute significantly to the atmospheric opacity on planet-wide scales; laboratory experiments and numerical models have demonstrated high levels of haze production in temperate and warm exoplanet atmospheres with $T_{\rm eq} \le 1000$ K \citep[e.g.,][]{fortney2013,morley2015,he2018,horst2018,kawashima2019,yu2021}.

Another related point of discrepancy between WASP-29b and WASP-80b is the amplitude of the water-vapor absorption feature at $\sim$1.4 $\mu$m: while WASP-80b's transmission spectrum displays a very prominent absorption feature, corresponding to a $4.2\sigma$ detection of H$_2$O, the WASP-29b spectrum shows no discernible feature. The so-called water amplitude $A_{H}$ quantifies the flatness of the transmission spectrum in the near-infrared and has become a commonly used metric for assessing broad trends in the impact of aerosol opacity on the growing ensemble of measured transmission spectra \citep[e.g.,][]{singstis,stevenson2016,crossfield2017,fu2017}. The most recent statistical analysis by \citet{dymont2021} carried out a multivariate search for trends among the body of available WFC3 transmission spectra for planets with $T_{\rm eq} < 1000$ K and uncovered significant covariances between $A_H$ and atmospheric scale height, planet gravity, and planet density.

Most of the published spectra used by \citet{dymont2021} were taken from the large population study by \citet{tsiaras2018}, which included WASP-29b and WASP-80b. We calculated new $A_H$ values for these planets based on our measured transmission spectra, following the techniques described in \citet{dymont2021}. After taking the weighted average of the transit depths measured in the two wavelength bins straddling 1.25 and 1.4 $\mu$m (Table~\ref{tab:specfit}), we obtained $A_H$ values of $-0.11 \pm 0.50$ and $1.00 \pm 0.52$ for WASP-29b and WASP-80b, respectively. Both of these results are consistent with the values in \citet{dymont2021}: $-0.17 \pm 0.47$ and $0.61 \pm 0.29$.


The aforementioned trends are reflected in a pairwise comparison of WASP-29b and WASP-80b. Namely, the attested positive correlation between $A_H$ and both $\log\,g_{p}$ and $\rho_{p}$, as well as the negative correlation between $A_H$ and $H$, are consistent with the difference in $A_H$ between the two planets. \citet{dymont2021} attributed the apparent reduction in haze opacity with increasing surface gravity to the corresponding increase in settling velocity, which works to remove aerosol particles from the upper atmosphere. Meanwhile, the relationship between $A_H$ and $H$ is more tentative, given the numerous caveats and simplifying assumptions behind the reported trend. For example, the assumption of a uniform atmospheric mean molecular weight ($\mu = 2.3$ amu) when computing $H$ disregards the broad range of retrieved atmospheric metallicities for temperate and warm gas giants. Ultimately, the production and distribution of aerosols are affected by a plethora of complex processes (e.g., vertical mixing and longitudinal transport; see, for example, \citealt{helling2008}, \citealt{parmentier2013}, and \citealt{gao2018etal}), which may relate to the fundamental properties of the planet in nontrivial ways.

In addition to H$_2$O, CO$_2$ was detected at moderately high ($3.4\sigma$) statistical significance in the transmission spectrum of WASP-80b, driven by the large relative transit-depth difference between the Spitzer 3.6 and 4.5 $\mu$m bandpasses. Notably, this is only the third significant detection of CO$_2$ from Spitzer transit observations, after WASP-17b \citep{alderson2022} and WASP-127b \citep{spake2021}. Meanwhile, there was no appreciable impact to the fit quality when we removed the opacity contributions of CO and CH$_4$. The retrieved limb temperature of WASP-80b ($\sim$800--1000 K at $1\sigma$) is notable because it lies in a region where the dominant atmospheric carbon reservoir is expected to transition from CO to CH$_4$ with decreasing temperature \citep[e.g.,][]{lodders2002,heng2016}. However, our retrieval results do not necessarily indicate that CO is absent or in low abundance along the terminator of WASP-80b. The absorption cross section of CO is significantly smaller than that of CO$_2$, so the opacity contribution of CO$_2$ dominates, even if CO is present at a higher mixing ratio. Indeed, the equilibrium abundance of CO calculated by the PLATON retrievals in the region of the atmosphere probed in transmission is higher than that of CO$_2$ by at least a factor of 5, even at the high-metallicity, low-temperature, low-C/O-ratio end of the $1\sigma$ confidence region. We also acknowledge that any interpretation of the relative CH$_4$, CO, and CO$_2$ abundances from our WASP-80b transmission spectrum hinges upon two broadband Spitzer points, which may themselves be unreliable, as demonstrated by the low Spitzer 4.5 $\mu$m transit depth that we obtained for WASP-29b (see also discussion in Section~\ref{subsec:wasp80retrieval} for WASP-80b).

Future high-resolution transmission spectra are needed to disentangle the opacity contributions of CO and CO$_2$ and definitively measure the mixing ratios of these major atmospheric species. As part of the JWST Guaranteed Time Observations, WASP-80b will be observed during both transit and secondary eclipse with the NIRCam instrument. These observations will provide uninterrupted wavelength coverage from 2.4 to 5 $\mu$m at a spectral resolution of $R = \lambda/\Delta\lambda > 1000$. This region contains the major absorption features of H$_2$O, CO, CO$_2$, and CH$_4$ (see Figure~\ref{wasp80retrieval}), allowing for detailed modeling of the chemical composition across the dayside hemisphere and limb. The precise molecular abundances will also support investigations into disequilibrium chemistry. Warm atmospheres such as that of WASP-80b are particularly susceptible to significant alterations from nonthermal processes, including vertical mixing, internal heating, photochemistry, and cold-trapping \citep[e.g.,][]{moses2013,zahnle2014,molaverdikhani2019,molaverdikhani2020,fortney2020}. The substantial improvements in wavelength coverage and precision afforded by JWST will enable us to move beyond the assumption of thermochemical equilibrium and more robustly explore the complexities of exoplanet atmospheres on both local and planet-wide scales.

\section{Summary}\label{sec:summary}

We have presented new 0.4--5.0 $\mu$m transmission spectra and 3.6 and 4.5 $\mu$m secondary eclipse measurements of two warm gas giants---WASP-29b and WASP-80b---using data collected by HST and Spitzer. The main findings of our study are summarized below:
\begin{enumerate}
\item The transmission spectrum of WASP-29b is flat and featureless throughout the optical and near-infrared, indicating significant opacity from large-particle aerosols along the day--night terminator. Other major atmospheric properties, such as limb temperature, C/O ratio, and metallicity, are poorly constrained in our retrievals.
\item WASP-80b shows a robust H$_2$O absorption feature at $\sim$1.4 $\mu$m, as well as a steep optical spectral slope. We retrieved a roughly solar C/O ratio of $0.50^{+0.14}_{-0.17}$. The attenuated amplitude of the water feature can be matched by both cloudy and cloud-free models with enhanced atmospheric metallicities ranging from several tens to a few hundred times solar, primarily driven by the large Spitzer 4.5 $\mu$m transit depth.. The cloud-free scenario requires unphysically high metallicities and is therefore disfavored. Meanwhile, the optical slope suggests the presence of fine-particle aerosols along the day--night terminator, although some amount of contamination from unocculted starspots may also be contributing, based on the stellar activity measured from long-term photometric monitoring of the target and X-ray emission observations.
\item The relatively deep Spitzer 4.5 $\mu$m transit depth of WASP-80b indicates strong molecular absorption from CO and/or CO$_2$. Upcoming higher-resolution transit spectroscopy with JWST will definitively resolve the transmission spectrum in this wavelength range and provide constraints of the individual molecular abundances.
\item From a global fit to all available Spitzer secondary eclipse light curves, we obtained 3.6 and 4.5 $\mu$m depths of $0.015\% \pm 0.008\%$ and $0.033\% \pm 0.006\%$ for WASP-29b, and $0.028\% \pm 0.005\%$ and $0.093^{+0.005}_{-0.006} \%$ for WASP-80b. The corresponding inferred dayside brightness temperatures are $937^{+45}_{-48}$ and $851^{+13}_{-14}$ K, respectively. These temperatures are comparable to the equilibrium temperatures calculated assuming zero Bond albedo and no heat recirculation to the nightside.
\end{enumerate}

\quad
\quad

This work is based on observations with the NASA/ESA Hubble Space Telescope, obtained at the Space Telescope Science Institute (STScI) operated by AURA, Inc. This work is also based in part on observations made with the Spitzer Space Telescope, which is operated by the Jet Propulsion Laboratory, California Institute of Technology under a contract with NASA. The research leading to these results has received funding from the European Research Council under the European Union's Seventh Framework Program (FP7/2007-2013)/ERC grant agreement no. 336792. Support for this work was also provided by NASA/STScI through grants linked to the HST-GO-12473 and HST-GO-14767 programs. Astronomy at Tennessee State University is supported by the State of Tennessee through its Centers of Excellence Program. I.W. is supported by an appointment to the NASA Postdoctoral Program at the NASA Goddard Space Flight Center, administered by the Universities Space Research Association under contract with NASA. J.S.F. acknowledges support from the Spanish State Research Agency projects AYA2016-79425-C3-2-P and PID2019-109522GB-C51. We also thank an anonymous referee for helpful comments that improved the manuscript.

\facilities{AIT, HST/STIS, HST/WFC3, Spitzer/IRAC, XMM-Newton.}
\software{\texttt{batman} \citep{batman}, \texttt{dynesty} \citep{speagle2020}, \texttt{emcee} \citep{emcee}, \texttt{ExoTEP} \citep{benneke2019,wong2020}, \texttt{LDTk} \citep{ldtk}.}


\appendix

\section{HST Spectroscopic Light Curves}
\restartappendixnumbering

\begin{figure*}[t]
\begin{center}
\includegraphics[width=14.6cm]{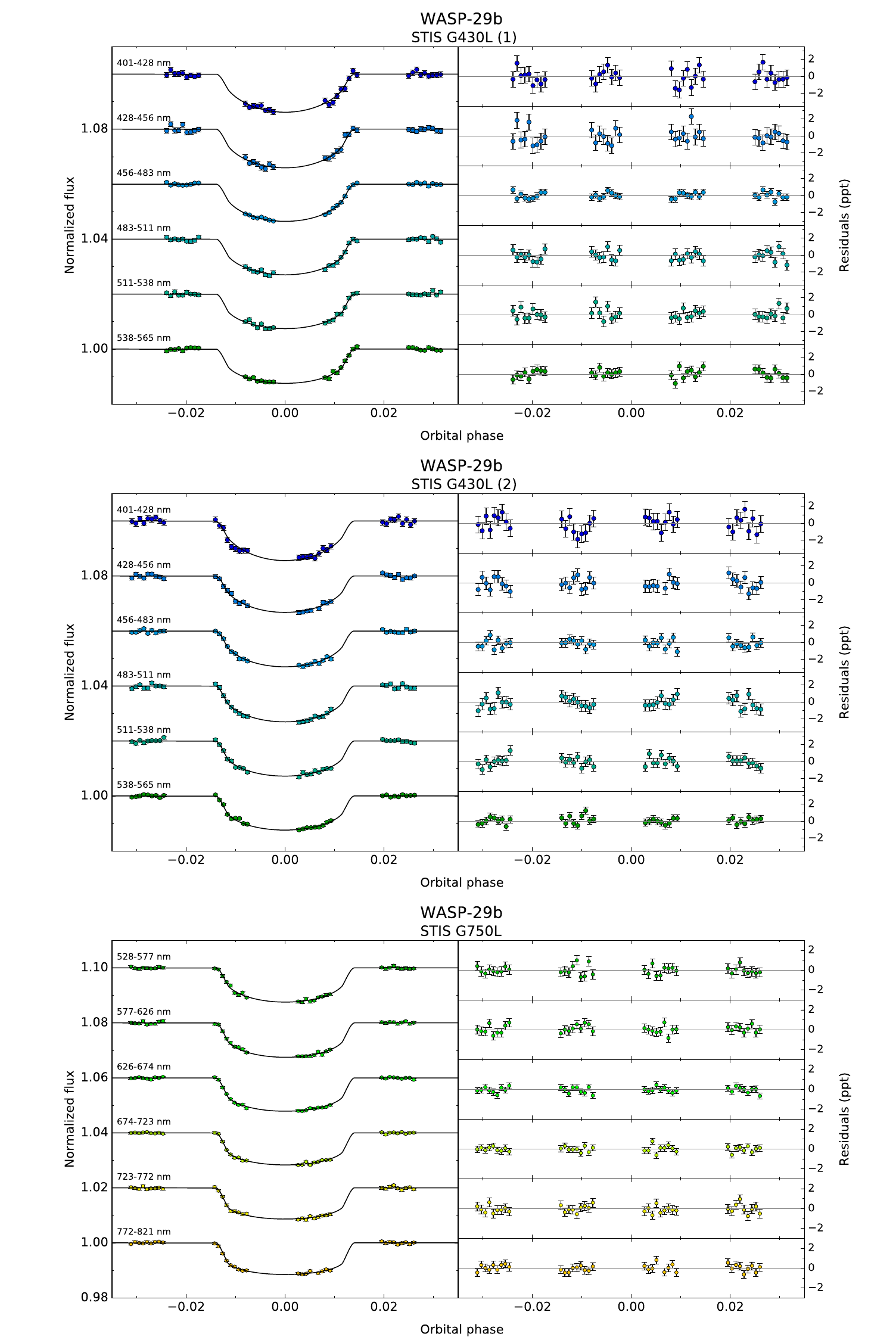}
\caption{A full compilation of systematics-corrected spectroscopic light curves (colored points) derived from the STIS G430L and STIS G750L observations of WASP-29b transits. The black curves show the best-fit transit models, while the corresponding residuals are plotted in the right panels. The light curves from the two STIS G430L visits are shown separately.}\label{fig:app1}
\end{center}
\end{figure*}

\begin{figure*}[t]
\begin{center}
\includegraphics[width=18cm]{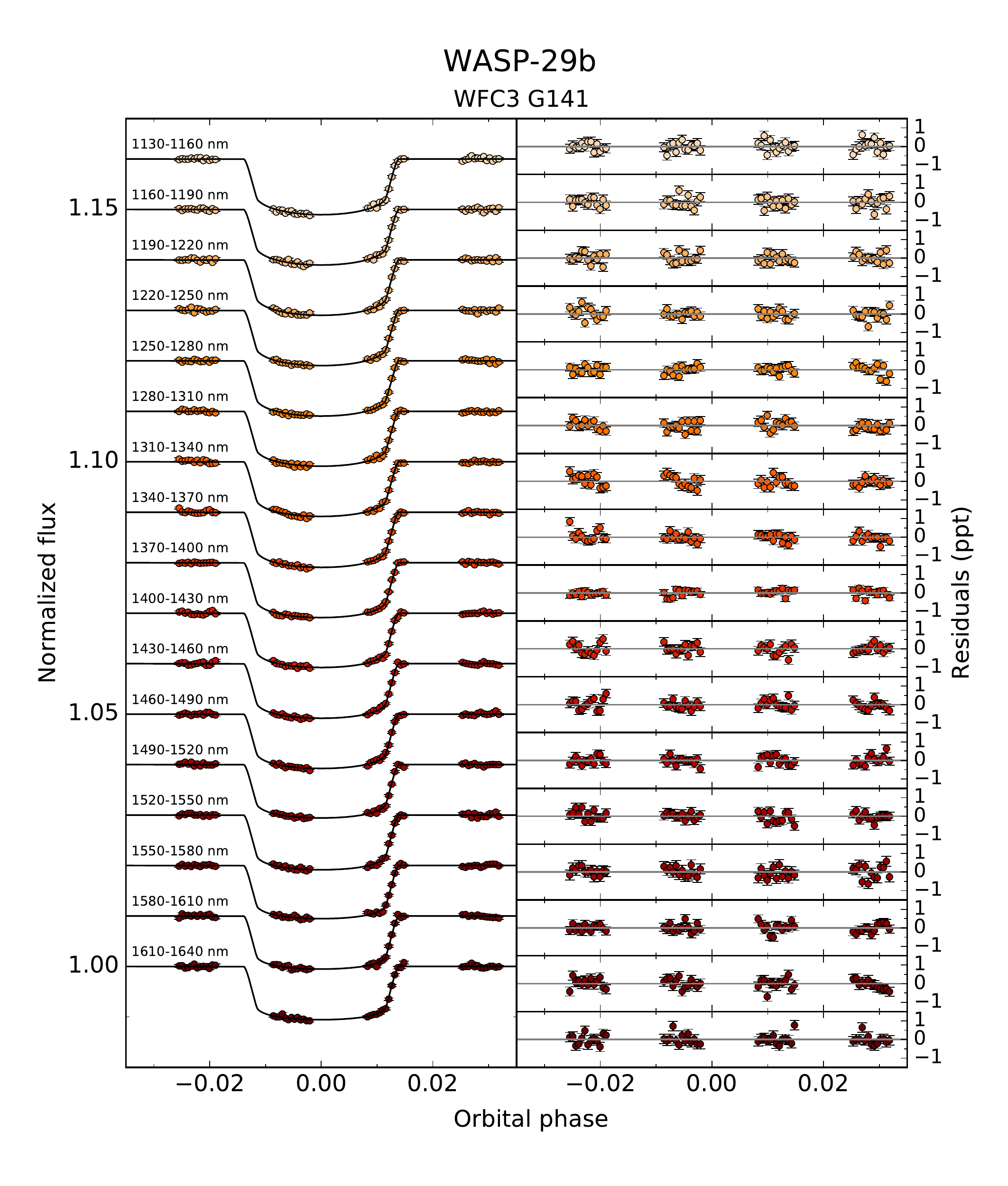}
\caption{Same as Figure~\ref{fig:app1}, but for the WFC3 G141 observation of WASP-29b.}\label{fig:app2}
\end{center}
\end{figure*}

\begin{figure*}[t]
\begin{center}
\includegraphics[width=15cm]{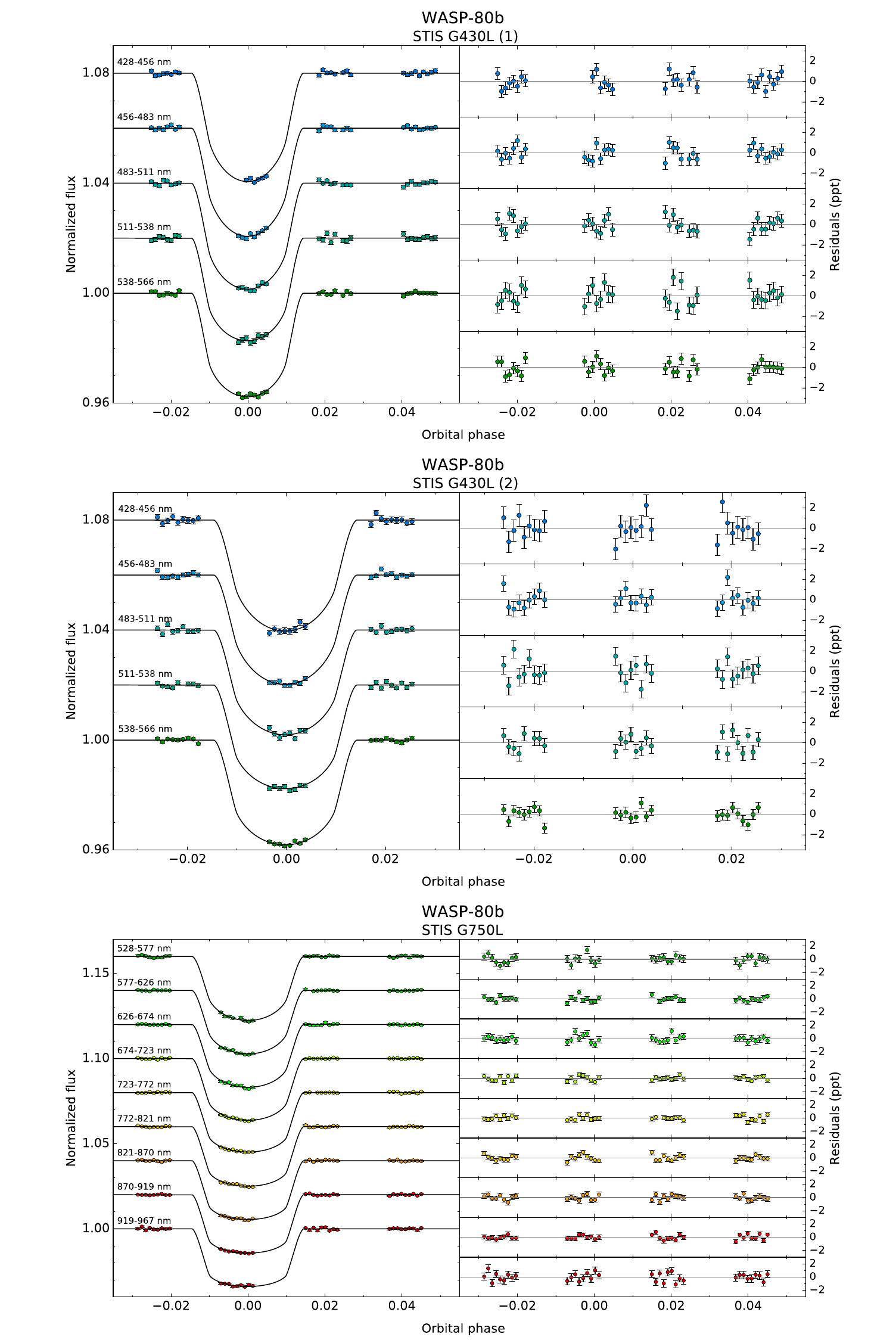}
\caption{Same as Figure~\ref{fig:app1}, but for the STIS observations of WASP-80b.}\label{fig:app3}
\end{center}
\end{figure*}

\begin{figure*}[t]
\begin{center}
\includegraphics[width=18cm]{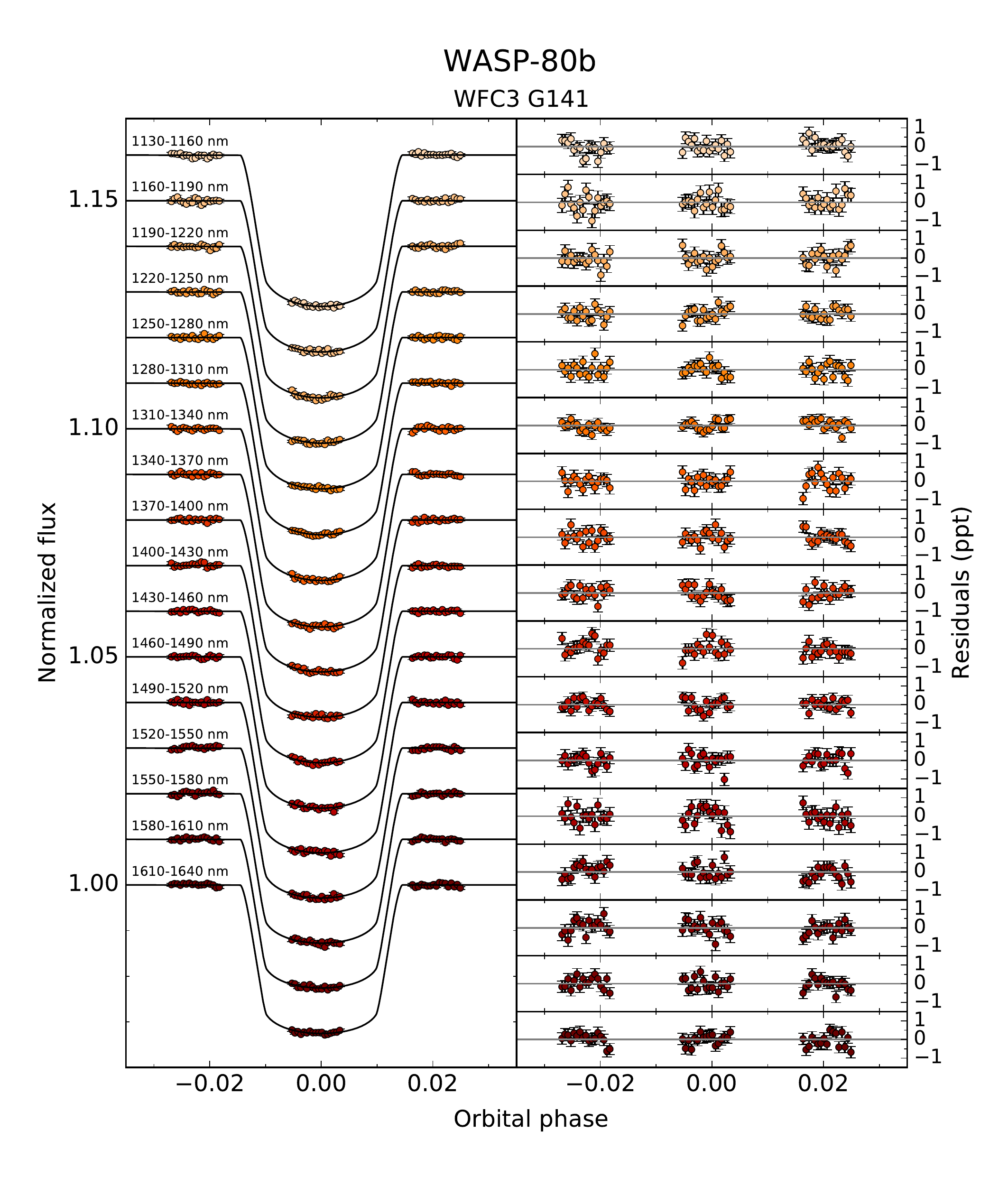}
\caption{Same as Figure~\ref{fig:app1}, but for the WFC3 G141 observation of WASP-80b.}\label{fig:app4}
\end{center}
\end{figure*}

Figures~\ref{fig:app1}--\ref{fig:app4} present a compilation of the HST spectroscopic light curves analyzed in this paper. The left panels show the systematics-corrected light curves for the wavelength bins listed in Table~\ref{tab:specfit}, excluding the narrowband spectroscopic light curves centered on the alkali absorption features. The corresponding best-fit transit models are overplotted in black. The right panels provide the residuals from the best-fit models.

\section{PLATON Retrieval Corner Plots}
\restartappendixnumbering

\begin{figure*}[b!]
\begin{center}
\includegraphics[width=\linewidth]{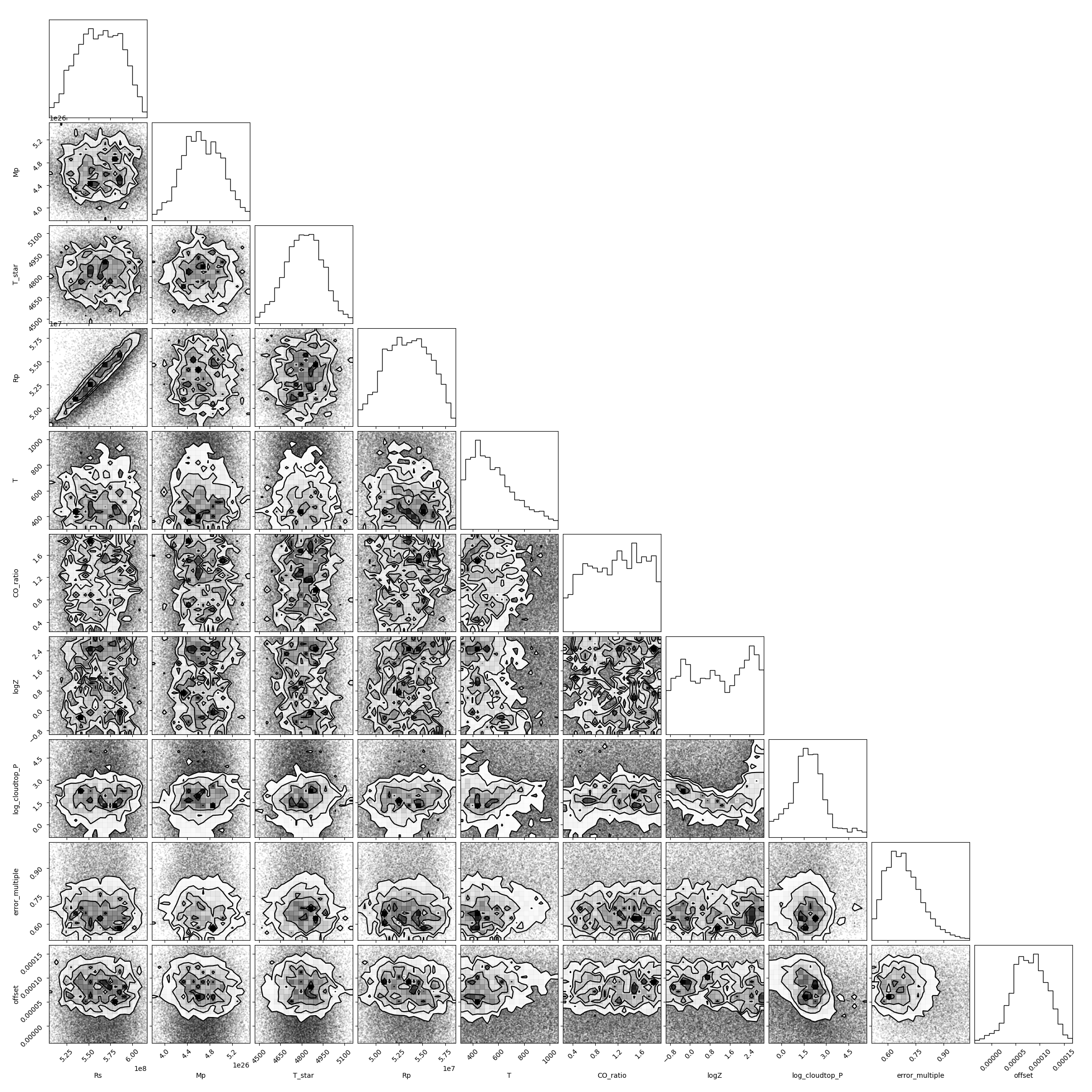}
\end{center}
\caption{Corner plot from the preferred PLATON atmospheric retrieval of the WASP-29b transmission spectrum, where the two Spitzer transit depths were excluded.} \label{wasp29corner}
\end{figure*}

\begin{figure*}[t!]
\begin{center}
\includegraphics[width=\linewidth]{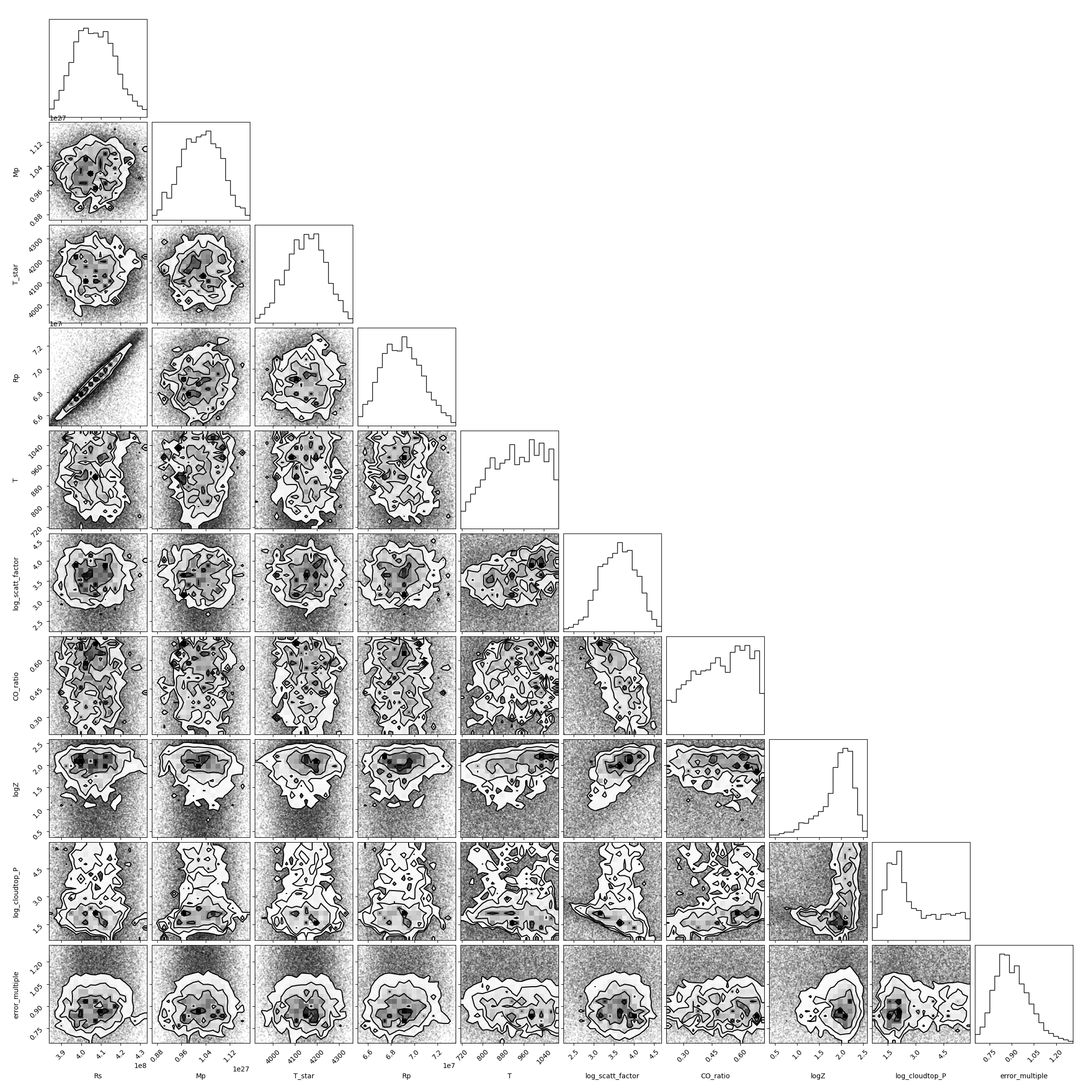}
\end{center}
\caption{Corner plot from the preferred PLATON atmospheric retrieval of the WASP-80b transmission spectrum.} \label{wasp80corner}
\end{figure*}

Figures~\ref{wasp29corner} and \ref{wasp80corner} show the corner plots from the preferred PLATON atmospheric retrieval runs for the WASP-29b and WASP-80b transmission spectra, respectively. The marginalized one-dimensional posteriors are shown along the diagonal. The contours in the two-dimensional posterior plots correspond to the 0.5$\sigma$, 1$\sigma$, 2$\sigma$, and 3$\sigma$ confidence regions.

\end{document}